\documentclass[%
 reprint,
 amsmath,amssymb,
 aps,
]{revtex4-2}

\usepackage{graphicx, float}% Include figure files
\usepackage{dcolumn}% Align table columns on decimal point
\usepackage{bm}% bold math
\usepackage{physics}
\usepackage{color}
\begin{document}

\title{Comparison of different exact generalized Langevin equations with a non-linear potential of mean force and an observable-dependent mass and friction}

\author{Benjamin J. A. Héry}
\affiliation{Department of Physics, Freie Universit\"at Berlin, 14195 Berlin, Germany.}
\author{Lucas Tepper}
\affiliation{Department of Physics, Freie Universit\"at Berlin, 14195 Berlin, Germany.}
\author{Roland R. Netz}
\email{corresponding author: rnetz@physik.fu-berlin.de}
\affiliation{Department of Physics, Freie Universit\"at Berlin, 14195 Berlin, Germany.}

\date{\today}

\begin{abstract}
    \setlength{\parindent}{0pt}
    The Mori-Zwanzig projection formalism constitutes a powerful and robust framework for 
    deriving equations of motion in terms of generalized Langevin equations (GLEs) for an 
    arbitrary observable using evolution and projection operators. Based on this 
    framework, we analyze the properties of four distinct GLEs for a scalar observable including a Markovian 
    force derived from a generally non-linear potential, a non-Markovian friction force, 
    and an orthogonal force, commonly interpreted as a random force. While all four GLEs 
    are exact, they differ in the memory friction kernel, which may either be dependent or 
    independent of the observable, and by the potential, which may either include or 
    exclude the effective kinetic energy of the observable. Inclusion of the kinetic energy in the potential is 
    advantageous for observables whose velocity satisfies Wick's theorem, since this reproduces the 
    correct distribution of the observable and its velocity even without contributions from the 
    friction force and the orthogonal force.
\end{abstract}

\maketitle

\section{\label{s1:introduction}Introduction}

Mori~\cite{mori_transport_1965}, Nakajima~\cite{nakajima_quantum_1958}, and
Zwanzig~\cite{zwanzig_ensemble_1960} introduced the operator formalism to derive 
generalized Langevin equations (GLEs), which are equations of motion for arbitrary coarse-grained reaction coordinates of complex systems~\cite{izvekov_microscopic_2013, 
izvekov_microscopic_2017, di_pasquale_systematic_2019, izvekov_microscopic_2019,
lee_multi-dimensional_2019, glatzel_comments_2021, klippenstein_introducing_2021, 
glatzel_interplay_2022, jung_non-markovian_2022, klippenstein_cross-correlation_2022, 
vroylandt_derivation_2022, vroylandt_position-dependent_2022, jung_dynamic_2023}. In the 
classical limit, the Mori-Zwanzig formalism is used to describe both equilibrium~\cite{izvekov_microscopic_2013, izvekov_microscopic_2017,
montoya-castillo_approximate_2017, ayaz_generalized_2022, vroylandt_derivation_2022, 
vroylandt_position-dependent_2022, ayaz_self-consistent_2022} and non-equilibrium 
phenomena~\cite{meyer_non-stationary_2017, izvekov_microscopic_2019,
vrugt_projection_2020, izvekov_mori-zwanzig_2021, jung_non-markovian_2022, 
netz_derivation_2024, hery_derivation_2024, izvekov_mori-zwanzig_2025}, with many 
applications in biology~\cite{hsu_zwanzig-mori_2009, glatzel_interplay_2022}, 
chemistry~\cite{freed_excluded_1976, lindenfeld_identity_1977, wertheimer_theory_1978, 
zippelius_kinetic_1978, schranner_dynamic_1979, wang_brillouin_1979, bose_time_1980, 
lanzafame_ultrafast_1992, schilling_mode_1997, chong_mode-coupling_1998, 
swinburne_phonon_2015, castellano_mode-coupling_2023}, computational 
physics~\cite{fulde_computation_1982, chorin_optimal_2000, li_variational_2007,
silbermann_coarse-grained_2008, darve_computing_2009, hijon_morizwanzig_2010,
kauzlaric_markovian_2013, mukhopadhyay_numerical_2013, carof_coarse_2014, carof_two_2014,
lang_tagged-particle_2014, parish_dynamic_2017, parish_non-markovian_2017, 
maeyama_extracting_2020, curtis_dynamic-mode_2021, klippenstein_introducing_2021, 
tian_data-driven_2021, wulkow_memory-based_2021, klippenstein_cross-correlation_2022, 
brunig_barrier-crossing_2022, brunig_pair-reaction_2022, brunig_time-dependent_2022, 
jung_dynamic_2023}, hydrodynamics~\cite{akcasu_theory_1970, duderstadt_calculation_1970, 
mashiyama_origin_1978, grossmann_correlation_1982, bixon_hard_1989, koide_transport_2008, 
gouasmi_priori_2017, bian_note_2018, kadam_dynamic_2022}, and statistical 
physics~\cite{nordholm_systematic_1975, tokuyama_statistical-mechanical_1976,
zwanzig_nonlinear_1978, kawasaki_projector_1992, vojta_charge_1998, lamba_variable-range_1999, givon_existence_2005, lyubimov_first-principle_2011,
venturi_mori-zwanzig_2017, zhu_generalized_2020, debets_generalized_2021,
te_vrugt_understanding_2022, fiorentino_green-kubo_2023}. Its strength lies in the exact 
nature of the resulting GLEs, which are derived without approximations or ad hoc 
assumptions and display a universal structure: a Markovian term that captures the 
force from a potential, a non-Markovian friction force, and an orthogonal 
force that is often treated as random noise~\cite{izvekov_mori-zwanzig_2017}.

In the present paper, we use different projection operators to derive and compare four different
GLEs for a scalar observable of interest $A$ that depends on positional degrees of freedom. We
concentrate on the special case when its corresponding velocity $\dot{A}$ satisfies Wick's theorem. Three
of the GLEs have been presented before~\cite{ayaz_self-consistent_2022, vroylandt_derivation_2022,
vroylandt_position-dependent_2022}. All four GLEs constitute effective equations of motion, with a
Markovian force that stems from a potential, a non-Markovian friction force that couples linearly to
$\dot{A}$ via a memory kernel, and an orthogonal force. While each GLE features the same
observable-dependent mass, two of their potentials include a kinetic energy contribution, and in two
of the GLEs the memory kernels are observable-dependent. In general, the given GLEs do not satisfy a
second moment relation (SMR) where the memory kernel is proportional to the second moment of the
orthogonal force. The inclusion of the kinetic energy terms renders the corresponding GLEs
particularly suitable for the analysis of complex data since in this case the exact joint
distribution of $A$ and $\dot{A}$ follows from the Boltzmann factor of the potential without
any contributions from the non-Markovian friction force or the orthogonal force.

We illustrate our results for the example of the folding dynamics of Villin from previous extensive MD simulations~\cite{lindorffLarsen_howFastFoldingProteins_2011}, described by the
fraction-of-native-contacts reaction coordinate $A(t)$~\cite{dalton_fast_2023}. Although the velocity of this
observable does not satisfy Wick's theorem exactly, the numerical results show that the GLEs
accurately approximate the joint distribution of $A(t)$ and its velocity.

\section{\label{s2:mori_zwanzig}Results}

\subsection*{\label{s2_ss1:generalities}Generic derivation of a GLE}

We consider a classical system of $N$ particles or atoms in three-dimensional space with the generalized
position vector $\vec{r} \equiv \left( r_{1}, \cdots, r_{3N} \right)^{T}$, the generalized momentum
vector $\vec{p} \equiv \left( p_{1}, \cdots, p_{3N} \right)^{T}$, and the microscopic state of the
system $\vec{w} \equiv \left( r_{1}, \cdots, r_{3N}, p_{1}, \cdots, p_{3N} \right)^{T}$, which
evolves in phase space denoted as $W = \mathbb{R}^{6N}$. We assume that the microscopic dynamics of
the system are derived from the generic time-independent many-body Hamiltonian
\begin{align}
    \label{s2_ss1_def:hamiltonian}
    & H(\vec{w}) \equiv \sum_{n = 1}^{3N} \frac{p_{n}^{2}}{2m_{n}} + V(\vec{r}),
\end{align}
where the first term represents the particle kinetic energy with $m_{n}$ the mass of the $n$-th coordinate
and $V(\vec{r})$ accounts for a generic interaction potential. The expectation value of an arbitrary
scalar time-independent Schrödinger-type observable $O_{S}(w)$ is given by
\begin{align}
    \label{s2_ss1_def:ensemble_expectation}
    & o(t) \equiv \int d^{6N}\vec{w} \: \rho(t, \vec{w}) O_{S}(\vec{w}).
\end{align}
For Hamiltonian dynamics, the time-dependent probability density distribution $\rho(t, \vec{w})$ is
determined by the Liouville equation
\begin{align}
    \label{s2_ss1_eq:liouville_equation}
    & \frac{\partial \rho(t, \vec{w})}{\partial t} = - \mathcal{L} \rho(t, \vec{w})
\end{align}
with the solution $\rho(t, \vec{w}) = e^{-t\mathcal{L}} \rho(0, \vec{w})$ and where
\begin{align}
    \label{s2_ss1_def:liouville_operator}
    & \mathcal{L} = \sum_{n = 1}^{3N} \left\lbrack \frac{p_{n}}{m_{n}} \frac{\partial}{\partial r_{n}} - \frac{\partial V(\vec{r})}{\partial r_{n}} \frac{\partial}{\partial p_{n}} \right\rbrack
\end{align}
defines the anti-self-adjoint Liouville operator. In sec.~\ref{si_1:generic_observable} of
the supplementary material, we show that the expectation value of $O_{S}(w)$ can alternatively be written as
\begin{align}
    \label{s2_ss1_eq:ensemble_expectation}
    & o(t) = \int d^{6N}\vec{w} \: \rho(0, \vec{w}) O(t, \vec{w}) \equiv \langle O(t, \vec{w}) \rangle,
\end{align}
where we defined the time-dependent Heisenberg-type observable
\begin{align}
    \label{s2_ss1_def:generic_observable}
    & O(t, \vec{w}) \equiv e^{t \mathcal{L}} O_{S}(\vec{w}).
\end{align}
In the following, we consider the observable
\begin{align}
    \label{s2_ss1_def:observable_of_interest}
    & A(t, \vec{w}) \equiv e^{t \mathcal{L}} A_{S}(\vec{r}),
\end{align}
where we choose $A_{S}(\vec{r})$ to depend only on positional degrees of freedom, and we review the
main steps to derive a GLE for $A(t, \vec{w})$. First, we compute the first and the second partial
time derivatives of $A(t, \vec{w})$,
\begin{align}
    \label{s2_ss1_eq:velocity_acceleration_observables}
    \left\{
    \begin{array}{c}
        \dot{A}(t, \vec{w}) = e^{t \mathcal{L}} \dot{A}_{S}(\vec{w}) \\
        \ddot{A}(t, \vec{w}) = e^{t \mathcal{L}} \ddot{A}_{S}(\vec{w}),
    \end{array}
    \right.
\end{align}
where we use the notation $\dot{A}_{S}(\vec{w}) \equiv \mathcal{L} A_{S}(\vec{r})$ and
$\ddot{A}_{S}(\vec{w}) \equiv \mathcal{L}^{2} A_{S}(\vec{r})$. We define $\mathcal{I}$ as the
identity operator and introduce pairs of projection operators, $\mathcal{P}_{1}$ and
$\mathcal{Q}_{1}$ and $\mathcal{P}_{2}$ and $\mathcal{Q}_{2}$, which satisfy $\mathcal{I} =
\mathcal{P}_{1} + \mathcal{Q}_{1}$ and $\mathcal{I} = \mathcal{P}_{2} + \mathcal{Q}_{2}$, so that we
can decompose $\ddot{A}(t, \vec{w})$ as
\begin{align}
    \label{s2_ss1_eq:first_decomposition}
    & \ddot{A}(t, \vec{w}) = e^{t \mathcal{L}} (\mathcal{P}_{1} + \mathcal{Q}_{1}) \ddot{A}_{S}(\vec{w}).
\end{align}
Next, we recall the Dyson decomposition identity~\cite{kawasaki_simple_1973,
zwanzig_nonequilibrium_2001}
\begin{align}
    \label{s2_ss1_eq:dyson_identity}
    & e^{t \mathcal{L}} = \int_{0}^{t} ds \: e^{s \mathcal{L}} \mathcal{P}_{2} \mathcal{L} e^{\left( t - s \right) \mathcal{Q}_{2} \mathcal{L}} + e^{t \mathcal{Q}_{2} \mathcal{L}},
\end{align}
and obtain the generic GLE
\begin{align}
    \label{s2_ss1_eq:gle}
    \ddot{A}(t, \vec{w}) & = F_{\text{eff}}(t, \vec{w}) \\
    & + \int_{0}^{t} ds \: F_{\text{f}}(s, t-s, \vec{w}) + F_{\mathcal{Q}}(t, \vec{w}), \nonumber
\end{align}
where we defined the effective force
\begin{align}
    \label{s2_ss1_def:effective_force}
    & F_{\text{eff}}(t, \vec{w}) \equiv e^{t \mathcal{L}} \mathcal{P}_{1} \ddot{A}_{S}(\vec{w}),
\end{align}
the orthogonal force
\begin{align}
    \label{s2_ss1_def:orthogonal_force}
    & F_{\mathcal{Q}}(t, \vec{w}) \equiv e^{t \mathcal{Q}_{2} \mathcal{L}} \mathcal{Q}_{1} \ddot{A}_{S}(\vec{w}),
\end{align}
and the non-Markovian friction force
\begin{align}
    \label{s2_ss1_def:memory_kernel}
    & F_{\text{f}}(s, t, \vec{w}) \equiv e^{s \mathcal{L}} \mathcal{P}_{2} \mathcal{L} F_{\mathcal{Q}}(t, \vec{w}).
\end{align}
As $\mathcal{P}_{1}$ and $\mathcal{P}_{2}$ are in general different projection operators, the procedure
leading to eq.~\ref{s2_ss1_eq:gle} is thus a dual-projection scheme~\cite{vroylandt_derivation_2022,ayaz_self-consistent_2022, kiefer_analysis_2025}.

\subsection*{\label{s2_ss2:list_projections}Different projection operators}

We choose the initial distribution $\rho(0, \vec{w})$ in eq.~\ref{s2_ss1_eq:ensemble_expectation} to be
the Boltzmann distribution determined by $H(\vec{w})$, i.e.
\begin{align}
    \label{s2_ss2_def:equilibrium_distribution}
    & \rho(0, \vec{w}) = \rho_{eq}(\vec{w}) \equiv \frac{e^{- \beta H(\vec{w})}}{Z(\beta)},
\end{align}
such that $\mathcal{L} \rho_{eq}(\vec{w}) = 0$ and where $Z(\beta) \equiv \int d^{6M}\vec{w}
\: e^{- \beta H(\vec{w})}$ is the partition function and $\beta \equiv (k_{B}T)^{-1}$ the
inverse thermal energy, and define the conditional ensemble expectation of the generic observable
$O(t, \vec{w})$ with respect to $A_{S}(\vec{r})$ by
\begin{align}
     \label{s2_ss2_def:conditional_expectation}
     & \langle O(t, \vec{w}) \rangle_{A} \equiv \frac{\langle O(t, \vec{w}) \delta ( A_{S}(\vec{r}) - A ) \rangle}{\langle \delta ( A_{S}(\vec{r}) - A ) \rangle},
\end{align}
In sec.~\ref{si_2:projection_formalism_discrete} and
sec.~\ref{si_3:projection_formalism_continuous}, we detail how to construct a projection operator, derive its
generic properties using the Mori-Zwanzig formalism, and introduce all projections that we require for the
different GLEs. In sec.~\ref{si_4:build_m}, we define the Mori projection
operator~\cite{mori_transport_1965}
\begin{align}
    \label{s2_ss2_def:mori_projection}
    & \mathcal{P}_{M} O(t, \vec{w}) \equiv \langle O(t, \hat{\vec{w}}) \rangle + \frac{\langle O(t, \hat{\vec{w}}) \dot{A}_{S}(\hat{\vec{w}}) \rangle}{\langle \dot{A}_{S}^{2}(\hat{\vec{w}}) \rangle} \dot{A}_{S}(\vec{w}) \\
    & + \frac{\langle O(t, \hat{\vec{w}}) ( A_{S}(\hat{\vec{r}}) - \langle A_{S}(\Tilde{\vec{r}}) \rangle ) \rangle}{\langle ( A_{S}(\hat{\vec{r}}) - \langle A_{S}(\Tilde{\vec{r}}) \rangle )^{2} \rangle} ( A_{S}(\vec{r}) - \langle A_{S}(\hat{\vec{r}}) \rangle ). \nonumber
\end{align}
where integration variables are denoted with a hat, i.e. $\hat{\vec{w}}$.

In sec.~\ref{si_5:build_h_z}, we define the positional Zwanzig projection
operator~\cite{izvekov_microscopic_2019, glatzel_interplay_2022}
\begin{align}
    \label{s2_ss2_def:zwanzig_projection}
    & \mathcal{P}_{Z} O(t, \vec{w}) \equiv \langle O(t, \hat{\vec{w}}) \rangle_{A_{S}(\vec{r})},
\end{align}
from which we construct the linear-velocity projection operator~\cite{vroylandt_derivation_2022}
\begin{align}
    \label{s2_ss2_def:hybrid_projection_z}
    \mathcal{P}_{LV} O(t, \vec{w}) & \equiv \mathcal{P}_{Z} O(t, \vec{w}) \\
    & + \frac{\langle O(t, \hat{\vec{w}}) \dot{A}_{S}(\hat{\vec{w}}) \rangle_{A_{S}(\vec{r})}}{\langle \dot{A}_{S}^{2}(\hat{\vec{w}}) \rangle_{A_{S}(\vec{r})}} \dot{A}_{S}(\vec{w}), \nonumber
\end{align}
which consists of $\mathcal{P}_{Z}$ plus an additional term that contains $\dot{A}_{S}(w)$. Finally, in sec.~\ref{si_6:build_h_z_z} we define the kinetic-energy projection
operator~\cite{ayaz_self-consistent_2022}
\begin{align}
    \label{s2_ss2_def:hybrid_projection_z_z}
    \mathcal{P}_{KE} O(t, \vec{w}) & \equiv \mathcal{P}_{LV} O(t, \vec{w}) \\
    & + \frac{\langle O(t, \hat{\vec{w}}) ( \dot{A}_{S}^{2}(\hat{\vec{w}}) - \langle \dot{A}_{S}^{2}(\tilde{\vec{w}}) \rangle_{A_{S}(\vec{r})} ) \rangle_{A_{S}(\vec{r})}}{\langle ( \dot{A}_{S}^{2}(\hat{\vec{w}}) - \langle \dot{A}_{S}^{2}(\tilde{\vec{w}}) \rangle_{A_{S}(\vec{r})} )^{2} \rangle_{A_{S}(\vec{r})}} \nonumber \\
    & \times \left( \dot{A}_{S}^{2}(\vec{w}) - \langle \dot{A}_{S}^{2}(\hat{\vec{w}}) \rangle_{A_{S}(\vec{r})} \right), \nonumber
\end{align}
which includes $\mathcal{P}_{LV}$ plus an additional term that contains $\dot{A}_{S}^{2}(\vec{w})$
and which can be interpreted as the effective kinetic energy of the observable.

\subsection*{\label{s2_ss3:final_form_gle}Four different GLEs}

By choosing different combinations of $\mathcal{P}_{M}$, $\mathcal{P}_{LV}$, $\mathcal{P}_{KE}$ for
$\mathcal{P}_{1}$ and $\mathcal{P}_{2}$ in the dual projection scheme, we derive four different
GLEs. The discussion in the main text focuses on observables $A(t, w)$ whose velocities satisfy the conditional
Wick's theorem given by
\begin{align}
    \label{s2_ss3_eq:wick_theorem}
    \left\{
    \begin{array}{c}
        \langle \dot{A}_{S}^{2n}(\hat{\vec{w}}) \rangle_{A} = \frac{(2n)!}{2^{n}n!} ( \langle \dot{A}_{S}^{2}(\hat{\vec{w}}) \rangle_{A} )^{n} \\
        \langle \dot{A}_{S}^{2n+1}(\hat{\vec{w}}) \rangle_{A} = 0
    \end{array}
    \right.
\end{align}
for integer $n$, in the supplementary material we also discuss the case where eq.~\ref{s2_ss3_eq:wick_theorem} does not hold.
First, by choosing $\mathcal{P}_{1} = \mathcal{P}_{LV}$ and $\mathcal{P}_{2} =
\mathcal{P}_{M}$, we obtain the observable-independent-friction GLE (I-GLE) \cite{vroylandt_derivation_2022}
\begin{align}
    \label{s2_ss3_eq:gle_ni_mk}
    \ddot{A}(t, \vec{w}) = & - \frac{1}{M_{\text{eff}}(A(t, \vec{w}))} \frac{d U_{\text{eff}}(A(t, \vec{w}))}{d A} \\ 
    & - \int_{0}^{t} ds \: \Gamma_{\text{I}}(t - s) \dot{A}(s, \vec{w}) + F_{\text{I}}(t, \vec{w}), \nonumber
\end{align}
where in sec.~\ref{si_7:markvoian_force_1} it is shown that the observable-dependent effective mass is defined by
\begin{align}
    \label{s2_ss3_def:effective_mass}
    % & \frac{1}{\beta M_{\text{eff}}(A)} \equiv \langle \dot{A}_{S}^{2}(\hat{w}) \rangle_{A},
    & M_{\text{eff}}(A) \equiv \frac{1}{\beta \langle \dot{A}_{S}^{2}(\hat{w}) \rangle_{A}},
\end{align}
while the effective potential is given by
\begin{align}
    \label{s2_ss3_def:effective_potential}
    & U_{\text{eff}}(A) \equiv U_{\text{pmf}}(A) + \frac{1}{\beta} \log M_{\text{eff}}(A),
\end{align}
and where the potential of mean force is given by
\begin{align}
    \label{s2_ss3_def:pmf}
    &  U_{\text{pmf}}(A) \equiv  - \frac{1}{\beta} \log \langle \delta ( A_{S}(\hat{\vec{r}}) - A ) \rangle.
\end{align}

Second, by choosing $\mathcal{P}_{1} = \mathcal{P}_{2} = \mathcal{P}_{LV}$, we obtain the
observable-dependent-friction GLE (D-GLE)\cite{vroylandt_position-dependent_2022}
\begin{align}
    \label{s2_ss3_eq:gle_ni_zk}
    \ddot{A}(t, \vec{w}) & = - \frac{1}{M_{\text{eff}}(A(t, \vec{w}))} \frac{d U_{\text{eff}}(A(t, \vec{w}))}{d A} \\
    & - \int_{0}^{t} ds \: \Gamma_{\text{D}}(t - s, A(s, \vec{w})) \dot{A}(s, \vec{w}) + F_{\text{D}}(t, \vec{w}). \nonumber
\end{align}

Third, by choosing $\mathcal{P}_{1} = \mathcal{P}_{KE}$ and $\mathcal{P}_{2} = \mathcal{P}_{M}$, we
obtain the observable-independent-friction kinetic-energy GLE (IKE-GLE)
\begin{align}
    \label{s2_ss3_eq:gle_i_mk}
    \ddot{A}(t, \vec{w}) = & - \frac{1}{M_{\text{eff}}(A(t, \vec{w}))} \frac{\partial U_{\text{eff}}^{KE}(A(t, \vec{w}), \dot{A}(t, \vec{w}))}{\partial A} \\
    & - \int_{0}^{t} ds \: \Gamma_{\text{IKE}}(t - s) \dot{A}(s, \vec{w}) + F_{\text{IKE}}(t, \vec{w}), \nonumber
\end{align}
where the effective potential is given by
\begin{align}
    \label{s2_ss3_def:effective_potential_kinetic_energy}
    U_{\text{eff}}^{KE}(A, \dot{A}) & = U_{\text{pmf}}(A) \\
    & - \frac{1}{2\beta} \log M_{\text{eff}}(A) + \frac{1}{2} M_{\text{eff}}(A) \dot{A}^{2}, \nonumber
\end{align}
see sec.~\ref{si_8:markvoian_force_2} for details.

Finally, by choosing $\mathcal{P}_{1} = \mathcal{P}_{KE}$ and $\mathcal{P}_{2} = \mathcal{P}_{LV}$,
we obtain the observable-dependent-friction kinetic-energy GLE (DKE-GLE) \cite{ayaz_self-consistent_2022}
\begin{align}
    \label{s2_ss3_eq:gle_i_zk}
    & \ddot{A}(t, \vec{w}) = - \frac{1}{M_{\text{eff}}(A(t, \vec{w}))} \frac{\partial U_{\text{eff}}^{KE}(A(t, \vec{w}), \dot{A}(t, \vec{w}))}{\partial A} \\
    & - \int_{0}^{t} ds \: \Gamma_{\text{DKE}}(t - s, A(s, \vec{w})) \dot{A}(s, \vec{w}) + F_{\text{DKE}}(t, \vec{w}). \nonumber
\end{align}

We show in sec.~\ref{si_9:kernel_1} that for the I-GLE and the IKE-GLE the observable-independent memory kernels are given by
\begin{align}
    \label{s2_ss3_def:kernel_1}
    & \Gamma_{\alpha}(t) \equiv \frac{\langle \ddot{A}_{S}\vec{w}) F_{\alpha}(t,\vec{w}) \rangle}{\langle \dot{A}_{S}^{2}\vec{w}) \rangle},
\end{align}
where $\alpha = I, IKE$ and $F_{\alpha}(t, \vec{w})$ satisfies
\begin{align}
    \label{s2_ss3_eq:orthogonal_force_properties}
    \left\{
    \begin{array}{c}
        \langle F_{\alpha}(t,\vec{w}) \rangle = 0 \\
        \langle F_{\alpha}(t,\vec{w}) A_{S}(\hat{\vec{r}}) \rangle = 0 \\
        \langle F_{\alpha}(t,\vec{w}) \dot{A}_{S}\vec{w}) \rangle = 0.
    \end{array}
    \right.
\end{align}
Furthermore, $F_{\alpha}(t, \vec{w})$ and $\Gamma_{\alpha}(t)$ satisfy a second-moment relation (SMR) if
$\langle F_{\alpha}(t, \hat{\vec{w}}) \rangle_{A} = 0$, i.e.
\begin{align}
    \label{s2_ss3_eq:SMR_relation}
    \Gamma_{\alpha}(t) = \frac{\langle F_{\alpha}(0, \vec{w}) F_{\alpha}(t, \vec{w}) \rangle}{\langle \dot{A}_{S}^{2}(\vec{w}) \rangle},
\end{align}
for $\alpha = I, IKE$.
We show in sec.~\ref{si_10:kernel_2} that the D-GLE and the DKE-GLE display
observable-dependent memory kernels given by
\begin{align}
    \label{s2_ss3_def:kernel_2}
   \Gamma_{\beta}(t, A) \equiv & \frac{\langle \ddot{A}_{S}(\hat{\vec{w}}) F_{\beta}(t, \hat{\vec{w}}) \rangle_{A}}{\langle \dot{A}_{S}^{2}(\hat{\vec{w}}) \rangle_{A}} \\
    & - \frac{d \log \langle \delta ( A_{S}(\hat{\vec{r}}) - A ) \rangle}{dA} \frac{\langle F_{\beta}(t, \hat{\vec{w}}) \dot{A}_{S}^{2}(\hat{\vec{w}}) \rangle_{A}}{\langle \dot{A}_{S}^{2}(\hat{\vec{w}}) \rangle_{A}} \nonumber \\
    & - \frac{1}{\langle \dot{A}_{S}^{2}(\hat{\vec{w}}) \rangle_{A}} \frac{\partial \langle F_{\beta}(t, \hat{\vec{w}}) \dot{A}_{S}^{2}(\hat{\vec{w}}) \rangle_{A}}{\partial A}, \nonumber
\end{align}
where $\beta = D, DKE$, and $F_{\beta}(t, \vec{w})$ satisfies
\begin{align}
    \label{s2_ss3_eq:orthogonal_force_properties_2}
    \left\{
    \begin{array}{c}
        \langle F_{\beta}(t, \hat{\vec{w}}) \rangle_{A} = 0 \\
        \langle F_{\beta}(t, \hat{\vec{w}}) A_{S}(\hat{\vec{r}}) \rangle_{A} = 0 \\
        \langle F_{\beta}(t, \hat{\vec{w}}) \dot{A}_{S}(\hat{\vec{w}}) \rangle_{A} = 0,
    \end{array}
    \right.
\end{align}
% TODO here Roland wants you to write out explicitly why this implies that eq. 32 holds
which implies that eq.~\ref{s2_ss3_eq:orthogonal_force_properties} holds for $F_{\beta}(t, \hat{\vec{w}})$.
Furthermore, we show that $F_{\beta}(t, \vec{w})$ and
$\Gamma_{\beta}(t)$ satisfy an observable-dependent SMR, i.e.
\begin{align}
    \label{s2_ss3_eq:SMR_relation_2}
    \Gamma_{\beta}(t, A) = \frac{\langle F_{\beta}(0, \hat{\vec{w}}) F_{\beta}(t, \hat{\vec{w}}) \rangle_{A}}{\langle \dot{A}_{S}^{2}(\hat{\vec{w}}) \rangle_{A}},
\end{align}
if $\langle F_{\beta}(t, \hat{\vec{w}}) \dot{A}_{S}^{2}(\hat{\vec{w}}) \rangle_{A} = 0$ or if $\langle F_{\beta}(t, \hat{\vec{w}}) \dot{A}_{S}^{2}(\hat{\vec{w}}) \rangle_{A}$ solves the partial differential equation (PDE) 
\begin{align}
    \label{s2_ss3_eq:pde_faa_correlation}
    \frac{\partial \langle F_{\beta}(t, \hat{\vec{w}}) \dot{A}_{S}^{2}(\hat{\vec{w}}) \rangle_{A}}{\partial A} = & \beta  \langle F_{\beta}(t, \hat{\vec{w}}) \dot{A}_{S}^{2}(\hat{\vec{w}}) \rangle_{A} \\
    & \frac{d}{dA} \left\lbrack U_{\text{pmf}}(A) - \frac{1}{2 \beta} \log M_{\text{eff}}(A) \right\rbrack. \nonumber
\end{align}
% Benjamin, check the following sentence again, I think there was a numbering mixup in Rolands comment
% Also Rolands comment did not mention the condition $\langle F_{\beta}(t, \hat{\vec{w}}) \dot{A}_{S}^{2}(\hat{\vec{w}}) \rangle_{A} = 0$
We stress that the SMRs in eq.~\ref{s2_ss3_eq:SMR_relation} and \ref{s2_ss3_eq:SMR_relation_2} do
not hold in general, but only when the restrictive condition $\langle F_{\alpha}(t, \hat{\vec{w}})
\rangle_{A} = 0$ or eq.~\ref{s2_ss3_eq:pde_faa_correlation} are satisfied.

\begin{figure*}[t]
    \centering
    \includegraphics[width=\textwidth]{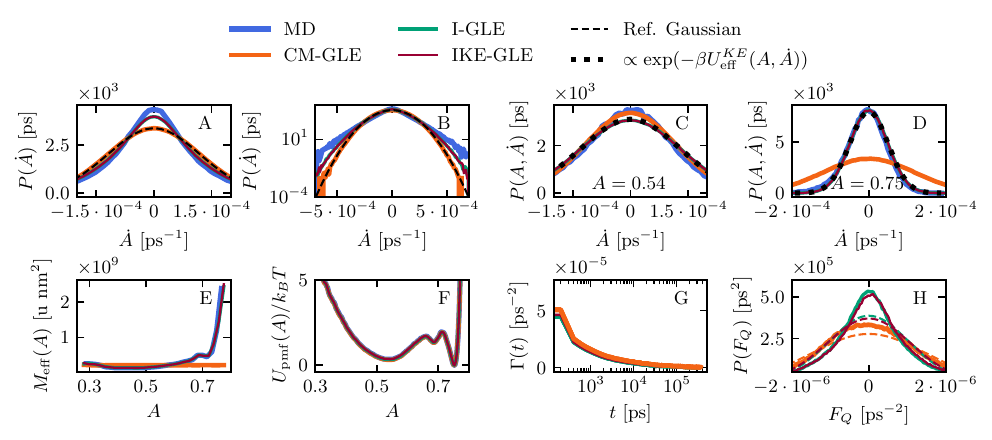}
    \caption{Comparison of MD simulation data for the protein Villin (blue) with different GLE simulation results in \textbf{A-F}
    and analysis of MD data using different GLEs in \textbf{G-H}. We use the observable-independent-friction GLE in eq.~\ref{s2_ss3_eq:gle_ni_mk} (I-GLE, green)
    the observable-independent-friction kinetic-energy IKE-GLE in eq.~\ref{s2_ss3_eq:gle_i_mk} (red) and the constant-mass CM-GLE in eq.~\ref{s2_ss4_eq:gle_cm} (orange).
    \textbf{A-B} Velocity distribution $P(\dot{A})$, where the dashed black lines indicate a Gaussian reference with zero mean and a standard deviation of $\sigma = 1 / (\beta
    M_{0} )$.
    Velocity distributions conditioned on $A = 0.54$ (\textbf{C}) and on $A = 0.75$ (\textbf{D}).
    The dotted black lines illustrate the advantage of the IKE-GLE by showing the Boltzmann distribution based on the potential
    $U_{\text{eff}}^{KE}(A, \dot{A})$ in eq.~\ref{s2_ss3_def:effective_potential_kinetic_energy}, which is proportional to $\exp \small( -\beta U_{\text{eff}}^{KE}(A, \dot{A}) \small)$.
    In \textbf{A-D}, results from the green and red lines are directly on top of each other.
    \textbf{E} The effective mass $M_{\text{eff}}(A)$ shows perfect agreement between the MD, I-GLE and IKE-GLE simulation data by the overlapping blue, green and red lines. 
    \textbf{F} The potentials $U_{\text{pmf}}(A)$ overlap exactly for all four simulation results.
    \textbf{G} The kernels $\Gamma_{\alpha}(t)$ (eq.~\ref{s2_ss3_def:kernel_1}) extracted from the MD data using the 
    I-GLE, IKE-GLE and CM-GLE are similar and lie on top of each other.
    \textbf{H} The random force distribution extracted using the three GLEs. The colored dashed lines indicate
    a Gaussian reference with zero mean and standard deviation of the corresponding data computed via $\sigma = \sqrt \langle ( F_{Q}(\hat{w}) - \langle F_{Q}(\Tilde{w}) \rangle )^2 \rangle$.}
    \label{s4_fig:comparison_gles}
\end{figure*}

An important distinction between the four GLEs is that the I-GLE in 
eq.~\ref{s2_ss3_eq:gle_ni_mk} and the D-GLE in eq.~\ref{s2_ss3_eq:gle_ni_zk} include 
$U_{\text{eff}}(A)$ in eq.~\ref{s2_ss3_def:effective_potential} while the IKE-GLE in 
eq.~\ref{s2_ss3_eq:gle_i_mk} and the DKE-GLE in eq.~\ref{s2_ss3_eq:gle_i_zk} feature
$U_{\text{eff}}^{KE}(A, \dot{A})$ defined in  
eq.~\ref{s2_ss3_def:effective_potential_kinetic_energy}. We note that the marginalization of the
Boltzmann distribution of $U_{\text{eff}}^{KE}(A, \dot{A})$ leads to
\begin{align}
    \label{s2_ss3_eq:potentials_relation}
    & \int_{-\infty}^{+\infty} d\dot{A} \: e^{- \beta U_{\text{eff}}^{KE}(A, \dot{A})} \propto e^{- \beta U_{\text{pmf}}(A)},
\end{align}
which demonstrates that the IKE-GLE and the DKE-GLE reproduce the expected marginal distribution of $A$
defined in eq.~\ref{s2_ss3_def:pmf} and the correct distribution for $\dot{A}$ without any contribution from the non-Markovian and orthogonal
forces (see sec.~\ref{si_11:cg_boltzmann} for details). This is a clear advantage of the GLE formulations including $U_{\text{eff}}^{KE}(A,
\dot{A})$ when the GLE is simulated by Markovian embedding.

\subsection*{\label{s2_ss4:numerics}Application to protein folding dynamics}

We apply our GLE formalism to previously published MD simulations of the folding of the protein
Villin~\cite{lindorffLarsen_howFastFoldingProteins_2011}. For this we use as reaction coordinate the
fraction of native contacts of Villin defined by
\begin{align}
    \label{s2_ss4_eq:q_coord}
    A(t) = \sum_{i = 1}^{N} \sum_{j > i} \frac{1}{1 + e^{\beta (s_{ij}(t) - \mu s_{ij}^{0})}},    
\end{align}
with $s_{ij}(t) = \Vert \vec{x}_{i}(t) - \vec{x}_{j}(t) \Vert$ \cite{dalton_fast_2023}, which
corresponds to an observable of interest of the general form
\begin{align}
    \label{s2_ss4_eq:choice_observable}
    A_{S}(\vec{r}) = \sum_{\alpha = 1}^{N_{d}} f_{\alpha} \left( D_{\alpha}(\vec{r}) \right),
\end{align}
where $D_{\alpha}(\vec{r}) \equiv \sum_{i_{\alpha}, j_{\alpha} = 1}^{N} \mu_{i_{\alpha}}
\mu_{j_{\alpha}} \vec{x}_{i_{\alpha}} \cdot \vec{x}_{j_{\alpha}}$ defines the bilinear
combination of the position vectors $\vec{x}_{i_{\alpha}} = \left( r_{3i_{\alpha} - 2},
r_{3i_{\alpha} - 1}, r_{3i_{\alpha}} \right)^{T}$, the $f_{\alpha} \left( D_{\alpha} \right)$ are arbitrary
functions, and $N_{d}$ is the number of functions considered. On one hand,
in sec.~\ref{si_12:example} we show that for $N_{d} = 1$, the observables given by
eq.~\ref{s2_ss4_eq:choice_observable} satisfy eq.~\ref{s2_ss3_eq:wick_theorem} and are described by an effective mass given by
\begin{align}
    \label{s2_ss4_eq:cg_mass}
    & M_{\text{eff}}(A) = \left( 4 f^{-1}(A)  \left( f'(f^{-1}(A)) \right)^{2} \sum_{i = 1}^{N} \frac{\mu_{i}^{2}}{m_{3i}} \right)^{-1},
\end{align}
where $f'(A)$ defines the functional inverse of $f(A)$ through $f^{-1}(f(A))$.
% TODO for Benjamin, I cite from Roland: (3B) under which condition is eq.~42 constant? say it!
On the other hand, we show in sec.~\ref{si_12:example} that for $N_{d} > 1$, observables 
of the form of eq.~\ref{s2_ss4_eq:choice_observable} do not generally fulfil
eq.~\ref{s2_ss3_eq:wick_theorem}. Since $A(t)$ in eq.~\ref{s2_ss4_eq:q_coord} can be represented in the form of  
eq.~\ref{s2_ss4_eq:choice_observable} by choosing 
\begin{align}
    \label{s2_ss4_eq:function}
    & f_{\alpha} \left( D \right) = \frac{1/N_{d}}{1 + \exp( \beta(\sqrt{D} - \mu s_{\alpha}^{0}) )}
\end{align}
and $N_{d} = 
\begin{pmatrix}
    N \\
    2 \end{pmatrix} > 1$ being the number of pairs formed by $N$ amino acids, we conclude that
% TODO Here, previously s2_ss4_def:potentials_equality was referenced, I think that was wrong and 
% changed it to eq 22 wiccks theorem
$A(t)$ in eq.~\ref{s2_ss4_eq:q_coord} does not satisfy eq.~\ref{s2_ss3_eq:wick_theorem}. In the following, we check
whether the GLEs derived in this paper can nevertheless be used approximately for the description of protein
folding using the non-linear reaction coordinate eq.~\ref{s2_ss4_eq:q_coord}.

In the following comparison, we also consider the constant-mass GLE (CM-GLE)
\begin{align}
    \label{s2_ss4_eq:gle_cm}
    \ddot{A}(t, \vec{w}) = & - \frac{1}{M_{0}} \frac{d U_{\text{eff}}(A(t, \vec{w}))}{d A} \\
    & - \int_{0}^{t} ds \: \Gamma_{\text{CM}}(t - s) \dot{A}(s, \vec{w}) + F_{\text{CM}}(t, \vec{w}), \nonumber
\end{align}
which follows from the I-GLE for a constant effective mass $M_{\text{eff}}(A) = M_{0} \equiv 1 / (\beta \langle
\dot{A}_{S}^{2}(\hat{\vec{w}}) \rangle)$. To simulate the CM-GLE, we can use the well-established
Markovian-embedding simulation method~\cite{ceriotti_colored-noise_2010, li_computing_2017,
ayaz_NonMarkovianModelingProtein_2021, tepper_accurate_2024} that maps the GLE to a set of Markovian
stochastic differential equations (SDEs). To simulate the I-GLE in eq.~\ref{s2_ss3_eq:gle_ni_mk} and the
IKE-GLE in eq.~\ref{s2_ss3_eq:gle_i_mk}, we use a recently introduced approximate Markovian
embedding~\cite{tepper_reactcoord_2026}. In plots \textbf{A-B}, we show that the distribution of
$\dot{A}$ from the MD data of Villin (blue lines) shows significant deviations from a reference
Gaussian distribution with $\langle \dot{A}^2 \rangle = 1 / (\beta M_0)$ (black dashed
line), which is expected based on the strong observable dependence of the effective mass
$M_{\mathrm{eff}}(A)$ in plot \textbf{E}. The embedding simulations for the I-GLE (green lines) and
IKE-GLE (red lines) exactly reproduce $M_{\mathrm{eff}}(A)$ in \textbf{E}, which suggests that the
slight differences visible between $P(\dot{A})$ for these GLEs and the MD data in \textbf{A-B}
reflect non-Gaussian effects. Plots
\textbf{C-D} show the joint probability density $P(A, \dot{A})$ for the two values $A = 0.54$, which
corresponds to the unfolded state within the potential $U_{\text{pmf}}(A)$ (see plot \textbf{F}) of
Villin, and $A = 0.75$, the folded state. The plots \textbf{C-D} demonstrate that $P(A, \dot{A})$ at
fixed $A$ is almost purely Gaussian and highlight that the inclusion of the effective mass in the I-GLE and
IKE-GLE (red and green lines) allows them to accurately reproduce the joint distribution $P(A, \dot{A})$, in contrast to the
CM-GLE (orange lines) in eq.~\ref{s2_ss4_eq:gle_cm} that is only able to reproduce $P(A)$.
The comparison with the Boltzmann factor of the effective energy $U_{\text{eff}}^{KE}(A, \dot{A})$
(dotted lines in plots \textbf{C,D}) demonstrates perfect agreement with the I-GLE and IKE-GLE simulation
results, which again points to non-Gaussian effects and reflects that eq.~\ref{s2_ss3_eq:wick_theorem} is
not satisfied for the reaction coordinate in eq.~\ref{s2_ss4_eq:q_coord}.
Plot \textbf{G} establishes that the memory kernels $\Gamma(t)$ extracted from the MD data of Villin
using Volterra equations~\cite{tepper_reactcoord_2026} based on all three GLEs are remarkably
similar.
% TODO maybe add short SI section with more details on extraction, actual kernel parameters 
% GLE simulation and fitting. and FQ extraction 
Given the extracted kernels, we compute the orthogonal forces $F_{Q}(t)$ by subtracting 
from the total acceleration at each frame of the MD trajectory the corresponding
potential terms of each GLE and the friction-kernel-integral term solved via numerical 
integration. In \textbf{H}, we plot the distribution of the corresponding orthogonal 
forces $P(F_{Q})$ for each of the three GLEs. By comparison with a Gaussian distribution 
that has zero mean and the same standard deviation as the obtained $F_{Q}(t)$ (denoted by the broken lines), we find 
$P(F_{Q})$ for the CM-GLE (orange line) to be approximately Gaussian, while a corresponding 
comparison for the I-GLE (green) and IKE-GLE (red) reveals a substantial non-Gaussian 
character. While $P(F_{Q})$ for the I-GLE and IKE-GLE are similar, the significant 
differences between $P(F_{Q})$ for these two GLEs and the CM-GLE signify that the 
inclusion of an observable-dependent $M_{\text{eff}}(A)$ strongly affects the non-
Markovian description of the underlying folding kinetics. 

The remarkable similarity between the GLE-simulation results for the I-GLE and IKE-GLE in plots 
\textbf{A-F} is not unexpected, as both use the exact same Markovian embedding, as detailed by 
Ref.~\cite{tepper_reactcoord_2026}.

In summary, Fig.~\ref{s4_fig:comparison_gles} establishes that the I-GLE in 
eq.~\ref{s2_ss3_eq:gle_ni_mk} and the IKE-GLE in eq.~\ref{s2_ss3_eq:gle_i_mk} are able to
reproduce MD data of proteins featuring strongly observable-dependent masses despite 
% TODO check that this references the correct si section, as Roland wanted
the fact that eq.~\ref{s2_ss3_eq:wick_theorem} is not strictly satisfied for the fraction
of native contacts of Villin, as shown in SI sec.~\ref{si_12:example}, which suggests
that the I-GLE and IKE-GLE formulations can be used as good approximations to describe protein
folding dynamics with the fraction of native contacts reaction coordinate in eq.~\ref{s2_ss4_eq:q_coord}.

\section{\label{s4:discussion }Discussion and conclusion}

In this paper, we present four exact GLEs and their respective derivations for a scalar observable of interest 
$A$ that satisfies Wick's theorem (cf. eq.~\ref{s2_ss3_eq:wick_theorem}) using different 
projection operators: the I-GLE, the D-GLE, the IKE-GLE, and the DKE-GLE. We highlight 
that when eq.~\ref{s2_ss3_eq:wick_theorem} holds, all four GLEs display the same effective 
mass $M_{\text{eff}}(A)$ but are generally characterized by different effective potentials and 
different memory kernels.

On one hand, we find that the effective potential of the IKE-GLE and the DKE-GLE 
exactly reproduces the joint distribution of $A$ and $\dot{A}$ without any contributions of the friction or orthogonal forces,
which, we argue, makes them preferable to GLEs that lack this property. Moreover,
sec.~\ref{si_8:markvoian_force_2} derives expressions for $M_{\text{eff}}(A)$ and 
$U_{\text{eff}}^{KE}(A, \dot{A})$ (cf. eqs~\ref{si_8_eq:initial_effective_force_2} 
and~\ref{si_8_eq:initial_effective_force_3}) when eq.~\ref{s2_ss3_eq:wick_theorem} 
does not hold and demonstrates that both quantities depend non-linearly on $U_{\text{pmf}}$. 
Therefore, satisfying Wick's theorem is a core prerequisite for the IKE-GLE and the 
DKE-GLE to correctly reproduce the joint statistics of $A$ and $\dot{A}$.

On the other hand, we show that no SMR between the orthogonal force and the memory kernel holds
in general for the four GLEs. As a matter of fact, in order for such a relation to hold the 
orthogonal force must satisfy additional conditions that go beyond eqs.~\ref{s2_ss3_eq:orthogonal_force_properties}
or~\ref{s2_ss3_eq:orthogonal_force_properties_2}. This is a subtle but important point, since
eqs.~\ref{s2_ss3_eq:orthogonal_force_properties} and~\ref{s2_ss3_eq:orthogonal_force_properties_2}
are direct results from the Mori-Zwanzig formalism. They specify which properties are expected to
hold for each orthogonal force based on the choice of $\mathcal{P}_{2}$ in Eq.~\ref{s2_ss1_def:orthogonal_force}.
Thus, the SMR is not expected to hold in the general case. This provides a reinterpretation of other
works that explain the breaking of the SMR as a consequence of a non-harmonic
$U_{\text{pmf}}(A)$~\cite{glatzel_interplay_2022, jung_limits_2026}.

Finally, Fig.~\ref{s4_fig:comparison_gles} compares MD-data for the fast-folding protein Villin
using the fraction of native contacts reaction coordinate $A(t)$ to simulations of the CM-GLE,
the I-GLE, and the IKE-GLE. We see that the I-GLE and the IKE-GLE
accurately sample the joint distribution $P(A, \dot{A})$ although the observable does not satisfy
eq.~\ref{s2_ss3_eq:wick_theorem}. This means that we can approximate the joint
probability distribution of $A(t)$ and its velocity by
% TODO, we removed this equation, so I just dropped the reference and s2_ss4_def:cg_boltzmann_distribution is now deep down in the SI, which is maybe not best referenced in the Conclusion. Check what you want to do here, Benjamin
% eqs.~\ref{s2_ss4_def:cg_boltzmann_distribution} and~\ref{s2_ss4_def:potentials_equality}, while using 
the corresponding Boltzmann distribution, while using 
the I-GLE or the IKE-GLE to accurately reproduce kinetic quantities of the MD simulations, including
the potential of mean force, the configuration-dependent mass, and the memory kernel,
which demonstrates the usefulness of the GLEs discussed in this paper.

\section{Acknowledgements}

We acknowledge support by Deutsche Forschungsgemeinschaft grant CRC 1449 "Dynamic Hydrogels at Biointerfaces", Project A02.

\newpage

\bibliography{sample}

\newpage

\appendix

\renewcommand{\thesection}{\arabic{section}} % Restore section numbering
\setcounter{section}{0} % Reset the section counter
\renewcommand{\appendixname}{} % Clear "Appendix" label

\newpage

% TODO, here you say Supplementary Material, but its different in the main text
%%%%%%%%%% Merge with Supplementary materials %%%%%%%%%%
\pagebreak
\newpage
\widetext
\begin{center}
    \textbf{\large Supplementary Material for: Comparison of different exact generalized Langevin equations with a non-linear potential of mean force and an observable-dependent mass and friction}\\
    \vspace{5mm}
    \textbf{Benjamin J. A. Hery, Lucas Tepper, Roland R. Netz}\\
    \vspace{1mm}
    \textit{Freie  Universität  Berlin,  Department  of  Physics,  Arnimallee  14,  14195  Berlin,  Germany}
\end{center}

%%%%%%%%%% Merge with Supplementary materials %%%%%%%%%%
%%%%%%%%%% Prefix a "S" to all equations, figures, tables and reset the counter %%%%%%%%%%
\setcounter{equation}{0}
\setcounter{figure}{0}
\setcounter{table}{0}
\setcounter{page}{1}
\makeatletter
\renewcommand{\thepage}{S\arabic{page}}
\renewcommand{\thesection}{S\arabic{section}}
\renewcommand{\thefigure}{S\arabic{figure}}
\renewcommand{\theequation}{S\arabic{section}.\arabic{equation}}

%\renewcommand{\bibnumfmt}[1]{[S#1]}
%\renewcommand{\citenumfont}[1]{S#1}
%%%%%%%%%% Prefix a "S" to all equations, figures, tables and reset the counter %%%%%%%%%%
%\section*{Supplementary Material}

\section{\label{si_1:generic_observable}From Schrödinger to Heisenberg observables}

We reformulate the expectation value $o(t)$ of the Schrödinger-type observable $O_{S}(\vec{w})$ in
eq.~\ref{s2_ss1_eq:ensemble_expectation} and derive the expression of the generic observable $O(t,
\vec{w})$ in eq.~\ref{s2_ss1_def:generic_observable} starting from the Liouville equation
eq.~\ref{s2_ss1_eq:liouville_equation} in the main text. First, we solve
eq.~\ref{s2_ss1_eq:liouville_equation} and obtain
\begin{align}
    \label{si_1_eq:probability_density}
    & \rho(t, \vec{w}) = e^{-t\mathcal{L}} \rho(0, \vec{w})
\end{align}
where $\rho(0, w)$ is the initial distribution and
\begin{align}
    \label{si_1_def:propagation_operator_backwards}
    & e^{-t\mathcal{L}} \equiv \sum_{n \geq 0} \frac{(-t)^{n}}{n!} \mathcal{L}^{n}
\end{align}
defines the evolution operator. We insert the expression of $\rho(t, w)$ from
eq.~\ref{si_1_def:propagation_operator_backwards} into the definition of $o(t)$ in
eq.~\ref{s2_ss1_def:ensemble_expectation} to obtain
\begin{align}
    \label{si_1_eq:ensemble_expectation}
    & o(t) = \int_{\mathbb{W}} d^{6N}\hat{\vec{w}} \: O_{S}(\hat{\vec{w}}) e^{-t \mathcal{L}} \rho(0, \hat{\vec{w}}).
\end{align}
Second, we use the definition of $e^{-t\mathcal{L}}$ in
eq.~\ref{si_1_def:propagation_operator_backwards} and the fact that $\mathcal{L}$ defines an anti-self-adjoint operator,
\begin{align}
    \label{si_1_def:anti_self_adjoindteness}
    & \int_{\mathbb{W}} d^{6N}\hat{\vec{w}}\: O_{1}(t, \hat{\vec{w}}) \mathcal{L} O_{2}(t', \hat{\vec{w}}) \equiv - \int_{\mathbb{W}} d^{6N}\hat{\vec{w}} \: O_{2}(t', \hat{\vec{w}}) \mathcal{L} O_{1}(t, \hat{\vec{w}}),
\end{align}
to recover both the expression of $O_{S}(\vec{w})$ in eq.~\ref{s2_ss1_eq:ensemble_expectation} and
the definition of $O(t, \vec{w})$ in eq.~\ref{s2_ss1_def:generic_observable}.

\section{\label{si_2:projection_formalism_discrete}Projection operator formalism for discrete sets}

We review the general theory of projection operator formalism presented by H. Vroylandt in
ref.~\cite{vroylandt_position-dependent_2022}. First, we consider a set
\begin{align}
    \label{si_2_def:projection_set_discrete}
    & \varepsilon \equiv \bigcup_{k \in I} \{ E_{k}(\vec{w}) \}
\end{align}
of observables where $I \subset \mathbb{N}$ is a set of indices. Thus, the projection operator
$\mathcal{P}_{\varepsilon}$ that is associated with $\varepsilon$ is defined by
\begin{align}
    \label{si_2_def:projection_operator_discrete}
    & \mathcal{P}_{\varepsilon} O(t, \vec{w}) \equiv \sum_{k, k' \in I} \langle O(t, \hat{\vec{w}}) E_{k}(\hat{\vec{w}}) \rangle (\hat{G}_{\varepsilon}^{-1})_{k, k'} E_{k'}(\vec{w})
\end{align}
where $\hat{G}_{\varepsilon}^{-1}$ is the inverse of the Gram matrix $\hat{G}_{\varepsilon}$ defined
component-wise by
\begin{align}
    \label{si_2_def:gram_matrix}
    & (\hat{G}_{\varepsilon})_{k, k'} \equiv \langle E_{k}(\hat{\vec{w}}) E_{k'}(\hat{\vec{w}}) \rangle
\end{align}
which satisfies
\begin{align}
    \label{si_2_eq:matricial_relation}
    & \sum_{k" \in I} (\hat{G}_{\varepsilon}^{-1})_{k, k"} (\hat{G}_{\varepsilon})_{k", k'} = \delta_{k, k'}.
\end{align}
Second, we show that $\mathcal{P}_{\varepsilon}$ defines a linear operator, i.e.
\begin{align}
    \label{si_2_eq:linearity}
    & \mathcal{P}_{\varepsilon} (\lambda_{1} O_{1}(t, \vec{w}) + \lambda_{2} O_{2}(t', \vec{w})) = \lambda_{1} \mathcal{P}_{\varepsilon} O_{1}(t, \vec{w}) + \lambda_{2} \mathcal{P}_{\varepsilon} O_{2}(t', \vec{w})
\end{align}
for any linear combination of Heisenberg-type observables $\lambda_{1} O_{1}(t, \vec{w}) +
\lambda_{2} O_{2}(t', \vec{w})$. We compute
\begin{align}
    \label{si_2_eq:linearity_2}
    & \mathcal{P}_{\varepsilon} (\lambda_{1} O_{1}(t, \vec{w}) + \lambda_{2} O_{2}(t', \vec{w})) = \sum_{k, k' \in I} \langle (\lambda_{1} O_{1}(t, \hat{\vec{w}}) + \lambda_{2} O_{2}(t', \hat{\vec{w}})) E_{k}(\hat{\vec{w}}) \rangle (\hat{G}_{\varepsilon}^{-1})_{k, k'} E_{k'}(\vec{w}), 
\end{align}
and use that $\langle \cdot \rangle$ defined in eq.~\ref{s2_ss1_eq:ensemble_expectation} and the
summations over the indices $k$ and $k'$ are linear operators to recover
eq.~\ref{si_2_eq:linearity}. Third, we show that each $E_{k}(\vec{w})$ is an invariant of
$\mathcal{P}_{\varepsilon}$, i.e.
\begin{align}
    \label{si_2_eq:invariant_quantities}
    & \mathcal{P}_{\varepsilon} E_{l}(\vec{w}) = E_{l}(\vec{w}).
\end{align}
We compute
\begin{align}
    \label{si_2_eq:invariant_quantities_2}
    & \mathcal{P}_{\varepsilon} E_{l}(\vec{w}) = \sum_{k, k' \in I} \langle E_{l}(\hat{\vec{w}}) E_{k}(\hat{\vec{w}}) \rangle (\hat{G}_{\varepsilon}^{-1})_{k, k'} E_{k'}(\vec{w}),
\end{align}
and use eq.~\ref{si_2_def:gram_matrix} and eq.~\ref{si_2_eq:matricial_relation} to recover
eq.~\ref{si_2_eq:invariant_quantities}. Fourth, we show that $\mathcal{P}_{\varepsilon}$ defines an
idempotent operator, i.e.
\begin{align}
    \label{si_2_eq:idempotency}
    & \mathcal{P}_{\varepsilon}^{2} O(t, \vec{w}) = \mathcal{P}_{\varepsilon} O(t, \vec{w}).
\end{align}
We compute
\begin{align}
    \label{si_2_eq:idempotency_2}
    & \mathcal{P}_{\varepsilon}^{2} O(t, \vec{w}) = \mathcal{P}_{\varepsilon} \left( \sum_{k, k' \in I} \langle O(t, \hat{\vec{w}}) E_{k}(\hat{\vec{w}}) \rangle (\hat{G}_{\varepsilon}^{-1})_{k, k'} E_{k'}(\vec{w}) \right)
\end{align}
and use that $\mathcal{P}_{\varepsilon}$ defines a linear operator and that each $E_{k}(\vec{w})$ is
an invariant of $\mathcal{P}_{\varepsilon}$ to recover eq.~\ref{si_2_eq:idempotency}. Fifth, we show
that $\mathcal{P}_{\varepsilon}$ defines a self-adjoint operator, i.e.
\begin{align}
    \label{si_2_eq:self_adjointedness}
    & \langle O_{1}(t, \hat{\vec{w}}) \mathcal{P}_{\varepsilon} O_{2}(t', \hat{\vec{w}}) \rangle = \langle O_{2}(t', \hat{\vec{w}}) \mathcal{P}_{\varepsilon} O_{1}(t, \hat{\vec{w}}) \rangle
\end{align}
for any Heisenberg-type observables $O_{1}(t, \vec{w})$ and $O_{2}(t', \vec{w})$. We compute
\begin{align}
    \label{si_2_eq:self_adjointedness_2}
    & \langle O_{1}(t, \hat{\vec{w}}) \mathcal{P}_{\varepsilon} O_{2}(t', \hat{\vec{w}}) \rangle = \sum_{k, k' \in I} \langle O_{1}(t, \hat{\vec{w}}) E_{k}(\hat{\vec{w}}) \rangle (\hat{G}_{\varepsilon}^{-1})_{k, k'} \langle O_{2}(t', \hat{\vec{w}}) E_{k'}(\hat{\vec{w}}) \rangle,
\end{align}
reorder the right-hand side of eq.~\ref{si_2_eq:self_adjointedness_2}, and use that
$\hat{G}_{\varepsilon}^{-1}$ is a symmetric positive real matrix to recover
eq.~\ref{si_2_eq:self_adjointedness}. Sixth, we define the complementary projection operator
\begin{align}
    \label{si_2_eq:complementary_projection_operator}
    & \mathcal{Q}_{\varepsilon} O(t, \vec{w}) \equiv (\mathcal{I}  - \mathcal{P}_{\varepsilon}) O(t, \vec{w})
\end{align}
where $\mathcal{I}O(t, \vec{w}) \equiv O(t, \vec{w})$ is the identity operator. By construction,
$\mathcal{Q}_{\varepsilon}$ also defines a linear, idempotent, and self-adjoint operator. Therefore,
for any generic observables $O_{1}(t, \vec{w})$ and $O_{2}(t', \vec{w})$, the correlation
\begin{align}
    \label{si_2_def:orthogonal_projection}
    & \langle \lbrack \mathcal{Q}_{\varepsilon} O_{1}(t, \hat{\vec{w}}) \rbrack \mathcal{P}_{\varepsilon} O_{2}(t', \hat{\vec{w}}) \rangle = 0
\end{align}
vanishes. In particular, it follows for each $E_{k}(\vec{w})$ that
\begin{align}
    \label{si_2_def:volterra_scheme}
    \langle E_{k}(\hat{\vec{w}}) \mathcal{Q}_{\varepsilon} O(t, \hat{\vec{w}}) \rangle = 0.
\end{align}
Seventh, we decompose $\mathcal{P}_{\varepsilon}$ into a sum of orthogonal projection operators, i.e.
\begin{align}
    \label{si_2_def:projection_operator_composition}
    \left\{
    \begin{array}{c}
        \mathcal{P}_{\varepsilon} O(t, \vec{w}) \equiv \underset{j \geq 1}{\sum} \mathcal{P}_{\Tilde{\varepsilon}_{j}} O(t, \vec{w}) \\
        \mathcal{P}_{\Tilde{\varepsilon}_{j}} O(t, \vec{w}) \equiv \langle O(t, \hat{\vec{w}}) \Tilde{E}_{l}(\hat{\vec{w}}) \rangle (\hat{\tilde{G}}_{\Tilde{\varepsilon}}^{-1})_{l, l} \Tilde{E}_{l}(\vec{w})
    \end{array}
    \right.
\end{align}
where
\begin{align}
    \label{si_2_eq:sub_projection_operator_orthogonality}
    \left\{
    \begin{array}{c}
        \mathcal{P}_{\Tilde{\varepsilon}_{j}} \mathcal{P}_{\Tilde{\varepsilon}_{j'}} O(t, \vec{w}) = \delta_{j, j'} \mathcal{P}_{\Tilde{\varepsilon}_{j}} O(t, \vec{w}) \\
        \langle \lbrack \mathcal{P}_{\Tilde{\varepsilon}_{j}} O_{1}(t, \hat{\vec{w}}) \rbrack \mathcal{P}_{\Tilde{\varepsilon}_{j'}} O_{2}(t', \hat{\vec{w}}) \rangle = \delta_{j, j'} \langle O_{1}(t, \hat{\vec{w}}) \mathcal{P}_{\Tilde{\varepsilon}_{j}} O_{2}(t', \hat{\vec{w}}) \rangle.
    \end{array}
    \right.
\end{align}
We use the fact that $\hat{G}_{\varepsilon}$ defines a symmetric real positive matrix so that there
exists an orthogonal matrix $\hat{T}_{\varepsilon \Tilde{\varepsilon}}$ such that
\begin{align}
    \label{si_2_eq:gram_matrix_diagonalization}
    & (\hat{\tilde{G}}_{\Tilde{\varepsilon}})_{l, l'} = \sum_{k, k' \in I} ( \hat{T}_{\varepsilon \Tilde{\varepsilon}}^{-1} )_{l, k} (\hat{G}_{\varepsilon})_{k, k'} ( \hat{T}_{\varepsilon \Tilde{\varepsilon}} )_{k', l'} = \delta_{l, l'}.
\end{align}
We use eq.~\ref{si_2_eq:gram_matrix_diagonalization} to compute the inverse of
$(\hat{G}_{\Tilde{\varepsilon}})_{l, l'}$ and plug the result into
eq.~\ref{si_2_def:projection_operator_discrete}. We recover
eq.~\ref{si_2_def:projection_operator_composition} where $\Tilde{E}_{l}(\vec{w})$ satisfies
\begin{align}
    \label{si_2_eq:transformation}
    & E_{k}(\vec{w}) = \sum_{l \in I} ( \hat{T}_{\varepsilon \Tilde{\varepsilon}} )_{k, l} \Tilde{E}_{l}(\vec{w}).
\end{align}
Therefore, decomposing a sum of orthogonal projection operators is equivalent to transforming
$\varepsilon$ into an orthogonal basis $\Tilde{\varepsilon}$. Since each
$\mathcal{P}_{\Tilde{\varepsilon}_{j}}$ defines the projection of one element of
$\Tilde{\varepsilon}$, eq.~\ref{si_2_eq:sub_projection_operator_orthogonality} follows naturally. In
practice, we diagonalize $\varepsilon$ using a Gram-Schmidt algorithm as for the Mori projection
operator $\mathcal{P}_{M}$ in sec.~\ref{si_4:build_m}.

\section{\label{si_3:projection_formalism_continuous}Projection operator formalism for continuous sets}

We extend the result of sec.~\ref{si_2:projection_formalism_discrete} to continuous projection sets of the form
\begin{align}
    \label{si_3_def:projection_set_continuous}
    & \varepsilon \equiv \bigcup_{z \in \mathbb{R}} \{ E(z, \vec{w}) \}.
\end{align}
Hence, the action of $\mathcal{P}_{\varepsilon}$ on $O(t, \vec{w})$ is now defined by
\begin{align}
    \label{si_3_def:projection_operator_continuous}
    & \mathcal{P}_{\varepsilon} O(t, \vec{w}) \equiv \int_{- \infty}^{+ \infty} dz \int_{- \infty}^{+ \infty} dz' \: \langle O(t, \hat{\vec{w}}) E(z, \hat{\vec{w}}) \rangle \mathcal{G}_{\varepsilon}^{-1}(z, z') E(z', \vec{w}),
\end{align}
where $\mathcal{G}_{\varepsilon}^{-1}$ is the inverse of the Gram kernel $\mathcal{G}_{\varepsilon}$ defined element-wise by
\begin{align}
    \label{si_3_def:gram_operator}
    & \mathcal{G}_{\varepsilon}(z, z') \equiv \langle E(z, \hat{\vec{w}}) E(z', \hat{\vec{w}}) \rangle
\end{align}
and satisfies
\begin{align}
    \label{si_3_eq:operator_relation}
    & \int_{- \infty}^{+ \infty} dz" \: \mathcal{G}_{\varepsilon}^{-1}(z, z") \mathcal{G}_{\varepsilon}(z", z') = \delta(z - z').
\end{align}
Similar to sec.~\ref{si_2:projection_formalism_discrete}, $\mathcal{P}_{\varepsilon}$ first defines a linear operator, i.e.
\begin{align}
    \label{si_3_eq:linearity}
    & \mathcal{P}_{\varepsilon} (\lambda_{1} O_{1}(t, \vec{w}) + \lambda_{2} O_{2}(t', \vec{w})) = \lambda_{1} \mathcal{P}_{\varepsilon} O_{1}(t, \vec{w}) + \lambda_{2} \mathcal{P}_{\varepsilon} O_{2}(t', \vec{w})
\end{align}
for any linear combination of Heisenberg-type observables $\lambda_{1} O_{1}(t, \vec{w}) +
\lambda_{2} O_{2}(t', \vec{w})$. Second, each $E(z, \vec{w})$ is invariant under
$\mathcal{P}_{\varepsilon}$, i.e.
\begin{align}
    \label{si_3_eq:operator_invariant}
    & \mathcal{P}_{\varepsilon} E(z", \vec{w}) = E(z", \vec{w}).
\end{align}
Third, $\mathcal{P}_{\varepsilon}$ defines an idempotent operator, i.e.
\begin{align}
    \label{si_3_eq:idempotency}
    & \mathcal{P}_{\varepsilon}^{2} O(t, \vec{w}) = \mathcal{P}_{\varepsilon} O(t, \vec{w}).
\end{align}
Fourth, $\mathcal{P}_{\varepsilon}$ defines a self-adjoint operator, i.e.
\begin{align}
    \label{si_3_eq:self_adjointedness}
    & \langle O_{1}(t, \hat{\vec{w}}) \mathcal{P}_{\varepsilon} O_{2}(t', \hat{\vec{w}}) \rangle = \langle O_{2}(t', \hat{\vec{w}}) \mathcal{P}_{\varepsilon} O_{1}(t, \hat{\vec{w}}) \rangle
\end{align}
for general Heisenberg-type observables $O_{1}(t, \vec{w})$ and $O_{2}(t', \vec{w})$. Fifth, the
complementary projection operator defined by
\begin{align}
    \label{si_3_eq:complementary_projection_operator}
    & \mathcal{Q}_{\varepsilon} O(t, \vec{w}) \equiv (\mathcal{I}  - \mathcal{P}_{\varepsilon}) O(t, \vec{w})
\end{align}
also defines a linear, idempotent, and self-adjoint operator. Therefore, we recover that for all
Heisenberg-type observables $O_{1}(t, \vec{w})$ and $O_{2}(t', \vec{w})$, the correlation
\begin{align}
    \label{si_3_def:orthogonal_projection}
    & \langle \lbrack \mathcal{Q}_{\varepsilon} O_{1}(t, \hat{\vec{w}}) \rbrack \mathcal{P}_{\varepsilon} O_{2}(t', \hat{\vec{w}}) \rangle = 0
\end{align}
vanishes. Hence, $\mathcal{P}_{\varepsilon}$ defines an orthogonal projection and we note in particular that
\begin{align}
    \label{si_3_def:volterra_scheme}
    & \langle E(z, \hat{\vec{w}}) \mathcal{Q}_{\varepsilon} O(t, \hat{\vec{w}}) \rangle = 0.
\end{align}
Sixth, we now consider the countable union of continuous sets
\begin{align}
    \label{si_3_eq:projection_set_continuous_1}
    & \varepsilon = \bigcup_{i \geq 1} \varepsilon_{i},
\end{align}
where each subset is defined by
\begin{align}
    \label{si_3_eq:projection_set_continuous_2}
    & \varepsilon_{i} = \bigcup_{z_{i} \in \mathbb{R}} \{ E_{i}(z_{i}, \vec{w}) \}.
\end{align}
Using these properties we obtain
\begin{align}
    \label{si_3_eq:projection_operator_continuous_1}
    & \mathcal{P}_{\varepsilon} O(t, \vec{w}) \equiv \sum_{i, i' \geq 1} \int_{- \infty}^{+ \infty} dz_{i} \int_{- \infty}^{+ \infty} dz_{i'}' \: \langle O(t, \hat{\vec{w}}) E_{i}(z_{i}, \hat{\vec{w}}) \rangle (\hat{\mathcal{G}}_{\varepsilon}^{-1}(z_{i}, z_{i'}'))_{i, i'} E_{i'}(z_{i'}', \vec{w}),
\end{align}
where $\hat{\mathcal{G}}_{\varepsilon}$ is a symmetric real positive Gram kernel matrix defined component-wise by
\begin{align}
    \label{si_3_eq:gram_operator_1}
    & (\hat{\mathcal{G}}_{\varepsilon}(z_{i}, z_{i'}'))_{i, i'} = \langle E_{i}(z_{i}, \hat{\vec{w}}) E_{i'}(z_{i'}', \hat{\vec{w}}) \rangle.
\end{align}
As in sec.~\ref{si_2:projection_formalism_discrete}, there exists an orthogonal kernel matrix $\hat{\mathcal{T}}_{\varepsilon \Tilde{\varepsilon}} (z', z')$ such that
\begin{align}
    \label{si_3_eq:gram_matrix_2}
    (\hat{\tilde{\mathcal{G}}}_{\Tilde{\varepsilon}}(z_{j}, z_{j'}'))_{j, j'} & = \sum_{i, i' \geq 1} \int_{- \infty}^{+ \infty} dz_{i} \int_{- \infty}^{+ \infty} dz_{i'}' \: ( \hat{\mathcal{T}}_{\varepsilon \Tilde{\varepsilon}}^{-1} (z_{j}, z_{i}))_{j, i} (\hat{\mathcal{G}}_{\varepsilon}(z_{i}, z_{i'}'))_{i, i'} ( \hat{\mathcal{T}}_{\varepsilon \Tilde{\varepsilon}} (z_{i'}', z_{j'}') )_{i' , j'} \\
    & = (\hat{\Tilde{\mathcal{G}}}_{\Tilde{\varepsilon}}(z_{j}, z_{j'}'))_{j, j'} \delta_{j, j'} \nonumber
\end{align}
so that $\mathcal{P}_{\varepsilon}$ can be decomposed into a sum of orthogonal projection operators, i.e.
\begin{align}
    \label{si_3_def:projection_operator_decomposition}
    \left\{
    \begin{array}{c}
        \mathcal{P}_{\varepsilon} O(t, \vec{w}) \equiv \underset{j \geq 1}{\sum} \mathcal{P}_{\Tilde{\varepsilon}_{j}} O(t, \vec{w}) \\
        \mathcal{P}_{\Tilde{\varepsilon}_{j}} O(t, \vec{w}) \equiv \int_{- \infty}^{+ \infty} dz \int_{- \infty}^{+ \infty} dz' \: \langle O(t, \hat{\vec{w}}) \Tilde{E}_{j}(z, \hat{\vec{w}}) \rangle (\hat{\tilde{\mathcal{G}}}_{\Tilde{\varepsilon}}^{-1}(z, z'))_{j, j} \Tilde{E}_{j'}(z', \vec{w})
    \end{array}
    \right.
\end{align}
such that
\begin{align}
    \label{si_3_def:sub_projection_operator_orthogonality}
    \left\{
    \begin{array}{c}
        \mathcal{P}_{\Tilde{\varepsilon}_{j}} \mathcal{P}_{\Tilde{\varepsilon}_{j'}} O(t, \vec{w}) = \delta_{j, j'} \mathcal{P}_{\Tilde{\varepsilon}_{j}} O(t, \vec{w}) \\
        \langle \lbrack \mathcal{P}_{\Tilde{\varepsilon}_{j}} O_{1}(t, \hat{\vec{w}}) \rbrack \mathcal{P}_{\Tilde{\varepsilon}_{j'}} O_{2}(t', \hat{\vec{w}}) \rangle = \delta_{j, j'} \langle O_{1}(t, \hat{\vec{w}}) \mathcal{P}_{\Tilde{\varepsilon}_{j}} O_{2}(t', \hat{\vec{w}}) \rangle,
    \end{array}
    \right.
\end{align}
and where $\Tilde{E}_{j}(z, \vec{w})$ is related to $E_{i}(z, \vec{w})$ via
\begin{align}
    \label{si_3_eq:transformation_1}
    & E_{i}(z_{i}, \vec{w}) = \sum_{j \geq 1} \int_{- \infty}^{+ \infty} dz_{j} \: (\hat{\mathcal{T}}_{\varepsilon \Tilde{\varepsilon}} (z_{i}, z_{j}))_{i, j} \Tilde{E}_{j}(z_{j}, \vec{w}).
\end{align}
In practice, this corresponds to transforming $\varepsilon$ into a countable union of orthogonal sets,
which we do by using a Gram-Schmidt algorithm when deriving the linear velocity projection operator
$\mathcal{P}_{LV}$ in sec.~\ref{si_5:build_h_z} and the action of the kinetic-energy projection
operator $\mathcal{P}_{KE}$ in sec.~\ref{si_6:build_h_z_z}.
 
\section{\label{si_4:build_m} Construction of $\mathcal{P}_{M}$}

We construct $\mathcal{P}_{M}$ in eq.~\ref{s2_ss2_def:mori_projection}. First, we consider the projection set
\begin{align}
    \label{si_4_def:projection_set}
    & \varepsilon_{M} \equiv \{1, A_{S}(\vec{r}), \dot{A}_{S}(\vec{w}) \}
\end{align}
and use the Gram-Schmidt algorithm to transform it into the orthogonal subset
\begin{align}
    \label{si_4_eq:tilde_projection_set}
    & \Tilde{\varepsilon}_{M} = \left\{1, A_{S}(\vec{r}) - \langle A_{S}(\hat{\vec{r}}) \rangle, \dot{A}_{S}(\vec{w}) - \langle \dot{A}_{S}(\hat{\vec{w}}) \rangle - \frac{\langle \dot{A}_{S}(\Tilde{\vec{w}}) (A_{S}(\hat{\vec{r}}) - \langle A_{S}(\Tilde{\vec{r}}) \rangle) \rangle}{\langle (A_{S}(\hat{\vec{r}}) - \langle A_{S}(\Tilde{\vec{r}}) \rangle)^{2} \rangle} (A_{S}(\vec{r}) - \langle A_{S}(\hat{\vec{r}}) \rangle) \right\}.
\end{align}
Hence, we introduce the linear mapping
\begin{align}
    \label{si_4_eq:linear_mapping}
    \left\{
    \begin{array}{c}
        1 \rightarrow 1 \\
        A_{S}(\vec{r}) \rightarrow A_{S}(\vec{r}) - a_{1} \\
        \dot{A}_{S}(\vec{w}) \rightarrow \dot{A}_{S}(\vec{w}) - a_{2} - a_{3} A_{S}(\vec{r})
    \end{array}
    \right.
\end{align}
and choose the values of $a_{1}$, $a_{2}$, and $a_{3}$ such that the new set $\Tilde{\varepsilon}_{M} = \{ 1, A_{S}(\vec{r}) - a_{1}, \dot{A}_{S}(w) - a_{2} - a_{3} A_{S}(\vec{r})\}$ is orthogonal. We obtain
\begin{align}
    \label{si_4_eq:linear_mapping_2}
    \left\{
    \begin{array}{c}
        a_{1} = \langle A_{S}(\hat{\vec{r}}) \rangle  \\
        a_{2} = \langle \dot{A}_{S}(\hat{\vec{w}}) \rangle - a_{3} \langle A_{S}(\hat{\vec{r}}) \rangle \\
        a_{3} = \frac{\langle \dot{A}_{S}(\hat{\vec{w}}) (A_{S}(\hat{\vec{r}}) - \langle A_{S}(\Tilde{\vec{r}}) \rangle ) \rangle}{\langle ( A_{S}(\hat{\vec{r}}) - \langle A_{S}(\Tilde{\vec{r}}) \rangle )^{2} \rangle}
    \end{array}
    \right..
\end{align}
Third, we recall that $\mathcal{L} \rho_{eq}(\vec{w}) = 0$, $\dot{A}_{S}(\vec{w}) = \mathcal{L} A_{S}(\vec{r})$ and $ A_{S}(\vec{r}) \dot{A}_{S}(\vec{w}) = \frac{1}{2} \mathcal{L} \left( A_{S}^{2}(\vec{r}) \right)^{2}$ to deduce that $\langle \dot{A}_{S}(\hat{\vec{w}}) \rangle = 0$ and $\langle \dot{A}_{S}(\hat{\vec{w}}) A_{S}(\hat{\vec{r}}) \rangle = 0$. Therefore,
\begin{align}
    \label{si_4_eq:projection_set_2}
    \Tilde{\varepsilon}_{M} = \{1, A_{S}(\vec{r}) - \langle A_{S}(\hat{\vec{r}}) \rangle, \dot{A}_{S}(\vec{w}) \}
\end{align}
where
\begin{align}
    \label{si_4_eq:gram_matrix_inverse}
    \hat{G}_{\Tilde{\varepsilon}_{M}}^{-1} = 
    \begin{pmatrix}
        1 & 0 & 0 \\
        0 & \frac{1}{\langle ( A_{S}(\hat{\vec{r}}) - \langle A_{S}(\Tilde{\vec{r}}) \rangle )^{2} \rangle} & 0 \\
        0 & 0 & \frac{1}{\langle \dot{A}_{S}^{2}(\hat{\vec{w}}) \rangle} 
    \end{pmatrix}.
\end{align}
Thus, using eq.~\ref{si_2_def:projection_operator_discrete}, where we substituted $\Tilde{\varepsilon}_{M}$ for $\Tilde{\varepsilon}$, we recover eq.~\ref{s2_ss2_def:mori_projection}, where we know from sec.~\ref{si_2:projection_formalism_discrete} that $\mathcal{P}_{M}$ defines a linear, idempotent and self-adjoint operator, that it defines an orthogonal projection, and that $1$, $A_{S}(\vec{r})$, and $\dot{A}_{S}(\vec{w})$ are its invariants.

\section{\label{si_5:build_h_z} Construction of $\mathcal{P}_{LV}$}

We construct $\mathcal{P}_{LV}$ in eq.~\ref{s2_ss2_def:hybrid_projection_z}. First, we consider the set
\begin{align}
    \label{si_5_def:projection_set}
    & \varepsilon_{LV} \equiv \varepsilon_{1} \cup \varepsilon_{\dot{A}},
\end{align}
where
% TODO should this k in R be z in R?
\begin{align}
    \label{si_5_def:projection_subsets}
    \left\{
    \begin{array}{c}
        \varepsilon_{1} \equiv \underset{k \in \mathbb{R}}{\bigcup} \{ \delta ( A_{S}(\vec{r}) - z ) \} \\
        \varepsilon_{\dot{A}} \equiv \underset{k \in \mathbb{R}}{\bigcup} \{ \dot{A}_{S}(\vec{w}) \delta ( A_{S}(\vec{r}) - z ) \}
    \end{array}
    \right..
\end{align}
According to eq.~\ref{si_3_eq:projection_operator_continuous_1}, the Gram kernel matrix
$\hat{\mathcal{G}}_{LV}$ associated with $\varepsilon_{LV}$ reads
\begin{align}
    \label{si_5_eq:gram_operator_blocks}
    \hat{\mathcal{G}}(z, z') & = 
    \begin{pmatrix}
        \mathcal{G}_{1, 1}(z, z') & \mathcal{G}_{1, \dot{A}}(z, z') \\
        \mathcal{G}_{\dot{A}, 1}(z, z') & \mathcal{G}_{\dot{A}, \dot{A}}(z, z')
    \end{pmatrix}
    \\
    & = \delta(z - z')
    \begin{pmatrix}
        \langle \delta ( A_{S}(\hat{\vec{r}} - z ) \rangle & \langle \dot{A}_{S}(\hat{\vec{w}}) \delta ( A_{S}(\hat{\vec{r}}) - z ) \rangle \\
        \langle \dot{A}_{S}(\hat{\vec{w}}) \delta ( A_{S}(\hat{\vec{r}}) - z ) \rangle & \langle \dot{A}_{S}^{2}(\hat{\vec{w}}) \delta ( A_{S}(\hat{\vec{r}}) - z ) \rangle
    \end{pmatrix}, \nonumber
\end{align}
where $\mathcal{G}_{1, 1}$ (resp. $\mathcal{G}_{\dot{A}, \dot{A}}$) quantifies the correlation
between observables that are both from the set $\varepsilon_{1}$, $\mathcal{G}_{\dot{A}, \dot{A}}$
quantifies the correlation between observables that are both from the set $\varepsilon_{\dot{A}}$,
and $\mathcal{G}_{1, \dot{A}}$ (or $\mathcal{G}_{\dot{A}, 1}$) quantifies the correlation between
observables from the set $\varepsilon_{1}$ and observables from the set $\varepsilon_{\dot{A}}$. We
compute
\begin{align}
    \label{si_5_eq:expectation_velocity_delta}
    \langle \dot{A}_{S}(\hat{\vec{w}}) \delta ( A_{S}(\hat{\vec{r}}) - z ) \rangle = \frac{1}{Z(\beta)} \sum_{j = 1}^{3N} \frac{1}{m_{j}} &  \left( \prod_{i = 1}^{3N} \int_{- \infty}^{+ \infty} d\hat{r}_{i} \: \delta ( A_{S}(\hat{\vec{r}}) - z ) \frac{\partial A_{S}(\hat{\vec{r}})}{\partial \hat{r}_{j}} e^{- \beta V(\hat{\vec{r}}) } \right) \\
    & \left( \prod_{i = 1}^{3N} \int_{- \infty}^{+ \infty} d\hat{p}_{i} \: \hat{p}_{j} e^{- \beta \overset{3N}{\underset{k = 1}{\sum}} \frac{\hat{p}_{k}^{2}}{2 m_{k}}} \right) \nonumber
\end{align}
and identify the integral over the components of $\vec{p}$ as the first moment of a multivariate
Gaussian integral that therefore vanishes, i.e. $\mathcal{G}_{1, \dot{A}}(z, z') =
\mathcal{G}_{\dot{A}, 1}(z, z') = 0$. Thus, $\varepsilon_{1}$ and $\varepsilon_{\dot{A}}$ are
orthogonal subsets and we can use eq.~\ref{si_3_def:projection_operator_decomposition} to decompose
$\mathcal{P}_{LV}$ into
\begin{align}
    \label{si_5_def:p_h_z}
    & \mathcal{P}_{LV} O(t, \vec{w}) \equiv \mathcal{P}_{1} O(t, \vec{w}) + \mathcal{P}_{\dot{A}} O(t, \vec{w}),
\end{align}
where $\mathcal{P}_{1}$ is the projection operator associated to $\varepsilon_{1}$ and
$\mathcal{P}_{\dot{A}}$ the projection operator associated to $\varepsilon_{\dot{A}}$. First, we
construct $\mathcal{P}_{1}$ and use eq.~\ref{si_3_eq:operator_relation} to compute
\begin{align}
    \label{si_5_eq:inverse_gram_operator_1}
    & \hat{\mathcal{G}}_{1}^{-1}(z, z') = \frac{\delta(z - z')}{\langle \delta ( A_{S}(\hat{\vec{r}}) - z' ) \rangle}.
\end{align}
We plug the result into eq.~\ref{si_3_def:sub_projection_operator_orthogonality}, where we
substituted $\delta ( A_{S}(\hat{\vec{r}}) - z )$ for $E(k, \vec{w})$, and use
eq.~\ref{s2_ss2_def:conditional_expectation} to obtain
\begin{align}
    \label{si_5_eq:p1_2}
    & \mathcal{P}_{1} O(t, \vec{w}) = \langle O(t, \hat{\vec{w}}) \rangle_{A_{S}(\vec{r})}.
\end{align}
Second, we construct $\mathcal{P}_{\dot{A}}$ and use eq.~\ref{si_3_eq:operator_relation} to compute
\begin{align}
    \label{si_5_eq:inverse_gram_operator_2}
    & \mathcal{G}_{\dot{A}}^{-1}(z, z') = \frac{\delta(z - z')}{\langle \dot{A}_{S}^{2}(\hat{\vec{w}}) \delta ( A_{S}(\hat{\vec{r}}) - z' ) \rangle}.
\end{align}
We plug the result in eq.~\ref{si_3_def:projection_operator_decomposition}, where we substituted
$\dot{A}_{S}(\vec{w}) \delta ( A_{S}(\hat{\vec{r}}) - z )$ for $E(k, \vec{w})$, and use
eq.~\ref{s2_ss2_def:conditional_expectation} to obtain
\begin{align}
    \label{si_5_eq:pdotA_2}
    & \mathcal{P}_{\dot{A}} O(t, \vec{w}) = \frac{\langle O(t, \hat{\vec{w}}) \dot{A}_{S}(\hat{\vec{w}}) \rangle_{A_{S}(\hat{\vec{r}})}}{\langle \dot{A}_{S}^{2}(\hat{\vec{w}}) \rangle_{A_{S}(\vec{r})}} \dot{A}_{S}(\vec{w}).
\end{align}
We plug the expressions of $\mathcal{P}_{1} O(t, \vec{w})$ and $\mathcal{P}_{\dot{A}} O(t, \vec{w})$
in eq.~\ref{si_5_def:p_h_z} and recover eq.~\ref{s2_ss2_def:hybrid_projection_z} where, according to
sec.~\ref{si_3:projection_formalism_continuous}, $\mathcal{P}_{LV}$ defines a linear, idempotent and
self-adjoint operator, defines an orthogonal projection, and has the following invariants: $\delta (
A_{S}(\vec{r}) - z )$ and $\delta ( A_{S}(\vec{r}) - z ) \dot{A}_{S}(\vec{w})$ $\forall k \in
\mathbb{R}$. Therefore, the observable 
% TODO is this \forall k \in \mathbb{R} correct?
\begin{align}
    \label{si_5_eq:invariant_observable_1}
    & f( A_{S}(\vec{r}) ) + \dot{A}_{S}(\vec{w}) g( A_{S}(\vec{r}) ) = \int_{- \infty}^{+ \infty} dz \: ( f(z) + g(z) \dot{A}_{S}(\vec{w}) ) \delta ( A_{S}(\vec{r}) - z )
\end{align}
is also an invariant of $\mathcal{P}_{LV}$, i.e.
\begin{align}
    \label{si_5_eq:invariant_observable_2}
    & \mathcal{P}_{LV} \left( f( A_{S}(\vec{r}) ) + \dot{A}_{S}(\vec{w}) g( A_{S}(\vec{r}) ) \right) = f( A_{S}(\vec{r}) ) + \dot{A}_{S}(\vec{w}) g( A_{S}(\vec{r}) ).
\end{align}

\section{\label{si_6:build_h_z_z} Construction of $\mathcal{P}_{KE}$}

We construct $\mathcal{P}_{KE}$ in eq.~\ref{s2_ss2_def:hybrid_projection_z_z}. First, we consider the set
\begin{align}
    \label{si_6_def:projection_set_1}
    & \varepsilon_{KE} \equiv \varepsilon_{1} \cup \varepsilon_{\dot{A}} \cup \varepsilon_{\dot{A}^{2}}
\end{align}
where
\begin{align}
    \label{si_6_def:projection_sub_sets_1}
    \left\{
    \begin{array}{c}
        \varepsilon_{1} \equiv \underset{k \in \mathbb{R}}{\bigcup} \{ \delta ( A_{S}(\vec{r}) - z ) \} \\
        \varepsilon_{\dot{A}} \equiv \underset{k \in \mathbb{R}}{\bigcup} \{ \dot{A}_{S}(\vec{w}) \delta ( A_{S}(\vec{r}) - z ) \} \\
        \varepsilon_{\dot{A}^{2}} \equiv \underset{k \in \mathbb{R}}{\bigcup} \{ \dot{A}_{S}^{2}(\vec{w}) \delta ( A_{S}(\vec{r}) - z ) \}
    \end{array}
    \right..
\end{align}
Second, we know from eq.~\ref{si_3:projection_formalism_continuous} that the associated Gram kernel matrix reads
\begin{align}
    \label{si_6_eq:gram_operator_matrix}
    \hat{\mathcal{G}}_{KE}(z, z') & =
    \begin{pmatrix}
        \mathcal{G}_{1, 1}(z, z') & \mathcal{G}_{1, \dot{A}}(z, z') & \mathcal{G}_{1, \dot{A}^{2}}(z, z') \\
        \mathcal{G}_{\dot{A}, 1}(z, z') & \mathcal{G}_{\dot{A}, \dot{A}}(z, z') & \mathcal{G}_{\dot{A}, \dot{A}^{2}}(z, z') \\
        \mathcal{G}_{\dot{A}^{2}, 1}(z, z') & \mathcal{G}_{\dot{A}^{2}, \dot{A}}(z, z') & \mathcal{G}_{\dot{A}^{2}, \dot{A}^{2}}(z, z')
    \end{pmatrix} \\
    & \equiv \delta(z - z')
    \begin{pmatrix}
        \langle \delta ( A_{S}(\hat{\vec{r}}) - z ) \rangle & 0 & \langle \dot{A}_{S}^{2}(\hat{\vec{w}}) \delta ( A_{S}(\hat{\vec{r}}) - z ) \rangle \\
        0 & \langle \dot{A}_{S}^{2}(\hat{\vec{w}}) \delta ( A_{S}(\hat{\vec{r}}) - z ) \rangle & \langle \dot{A}_{S}^{3}(\hat{\vec{w}}) \delta ( A_{S}(\hat{\vec{r}}) - z ) \rangle \\
        \langle \dot{A}_{S}^{2}(\hat{\vec{w}}) \delta ( A_{S}(\hat{\vec{r}}) - z ) \rangle & \langle \dot{A}_{S}^{3}(\hat{\vec{w}}) \delta ( A_{S}(\hat{\vec{r}}) - z ) \rangle & \langle \dot{A}_{S}^{4}(\hat{\vec{w}}) \delta ( A_{S}(\hat{\vec{r}}) - z ) \rangle
    \end{pmatrix} \nonumber
\end{align}
% TODO another question mark comment or squiggly line
where $\hat{\mathcal{G}}_{1, 1}(z, z')$, $\hat{\mathcal{G}}_{\dot{A}, \dot{A}}$, $\mathcal{G}_{1, \dot{A}}(z, z')$ and $\mathcal{G}_{\dot{A}, 1}(z, z')$ are the same functions as in sec.~\ref{si_5:build_h_z}, $\mathcal{G}_{\dot{A}^{2}, \dot{A}^{2}}$ denotes the correlation between observables of $\varepsilon_{\dot{A}^{2}}$, and $\mathcal{G}_{1, \dot{A}^{2}}$ (resp. $\hat{\mathcal{G}}_{\dot{A}, \dot{A}^{2}}(z, z')$) denotes correlations between observables from $\varepsilon_{1}$ and $\varepsilon_{\dot{A}^{2}}$ (resp. $\varepsilon_{\dot{A}}$ and $\varepsilon_{\dot{A}^{2}}$). We compute
\begin{align}
    \label{si_6_eq:gram_operator_sixth_block}
    \langle \dot{A}_{S}^{3}(\hat{\vec{w}}) \delta ( A_{S}(\hat{\vec{r}}) - z ) \rangle = \frac{1}{Z(\beta)} \sum_{j, j', j" = 1}^{3N} \frac{1}{m_{j} m_{j'} m_{j"}} & \left( \prod_{i = 1}^{3N} \int_{- \infty}^{+ \infty} d\hat{p}_{i} \: \hat{p}_{j} \hat{p}_{j'} \hat{p}_{j"} e^{- \beta \sum_{k = 1}^{3N} \frac{\hat{p}_{k}^{2}}{2 m_{k}}} \right) \\
    & \left( \prod_{i = 1}^{3N} \int_{- \infty}^{+ \infty} d\hat{r}_{i} \: \delta ( A_{S}(\hat{\vec{r}}) - z ) \frac{\partial A_{S}(\hat{\vec{r}})}{\partial \hat{r}_{j}} \frac{\partial A_{S}(\hat{\vec{r}})}{\partial \hat{r}_{j'}} \frac{\partial A_{S}(\hat{\vec{r}})}{\partial \hat{r}_{j"}} e^{- \beta V(\hat{\vec{r}}) } \right) \nonumber
\end{align}
and identify the integral over the components of $\vec{p}$ as the third moment of a multivariate
Gaussian integral that therefore vanishes, i.e. $\mathcal{G}_{\dot{A}, \dot{A}^{2}}(z, z') =
\mathcal{G}_{\dot{A}^{2}, \dot{A}}(z, z') = 0$. Thus, $\varepsilon_{1}$ and
$\varepsilon_{\dot{A}^{2}}$ are orthogonal to $\varepsilon_{\dot{A}}$ but are not orthogonal with
respect to each other. Third, we use the Gram-Schmidt algorithm and introduce the linear mapping
\begin{align}
    \label{si_6_eq:linear_mapping}
    \left\{
    \begin{array}{c}
        \delta ( A_{S}(\vec{r}) - z ) \rightarrow \delta ( A_{S}(\vec{r}) - z ) \\
        \dot{A}_{S}(\vec{w}) \delta ( A_{S}(\vec{r}) - z ) \rightarrow \dot{A}_{S}(\vec{w}) \delta ( A_{S}(\vec{r}) - z ) \\
        \dot{A}_{S}^{2}(\vec{w}) \delta ( A_{S}(\vec{r}) - z ) \rightarrow \dot{A}_{S}^{2}(\vec{w}) \delta ( A_{S}(\vec{r}) - z ) - \int_{- \infty}^{+ \infty} d\Tilde{z} \: \hat{\mathcal{T}}_{\varepsilon \Tilde{\varepsilon}}^{-1}(z, \Tilde{z}) \delta ( A_{S}(\vec{r}) - \Tilde{z} )
    \end{array}
    \right..
\end{align}
We choose $\hat{\mathcal{T}}_{\varepsilon \Tilde{\varepsilon}}^{-1}(z, \Tilde{z})$ such that
\begin{align}
    \label{si_6_eq:orthogonality_condition_1}
    & \langle \dot{A}_{S}^{2}(\hat{\vec{w}}) \delta ( A_{S}(\hat{\vec{r}}) - z ) \delta ( A_{S}(\hat{\vec{r}}) - \Bar{z} ) \rangle - \int_{- \infty}^{+ \infty} d\Tilde{z} \: \hat{\mathcal{T}}_{\varepsilon \Tilde{\varepsilon}}^{-1}(z, \Tilde{z}) \langle \delta ( A_{S}(\hat{\vec{r}}) - \Tilde{z} ) \delta ( A_{S}(\hat{\vec{r}}) - \Bar{z} ) \rangle = 0
\end{align}
and obtain $\hat{\mathcal{T}}_{\varepsilon \Tilde{\varepsilon}}^{-1}(z, \Tilde{z}) = \langle \dot{A}_{S}^{2}(\hat{\vec{w}}) \rangle_{\Bar{z}} \delta ( z - \Bar{z} )$. Thus, the modified set reads
\begin{align}
    \label{si_6_def:projection_set_2}
    & \Tilde{\varepsilon}_{KE} \equiv \varepsilon_{1} \cup \varepsilon_{\dot{A}} \cup \Tilde{\varepsilon}_{\dot{A}^{2}}
\end{align}
where
\begin{align}
    \label{si_6_def:projection_sub_sets_2}
    \left\{
    \begin{array}{c}
        \varepsilon_{1} \equiv \underset{k \in \mathbb{R}}{\bigcup} \{ \delta ( A_{S}(\vec{r}) - z ) \} \\
        \varepsilon_{\dot{A}} \equiv \underset{k \in \mathbb{R}}{\bigcup} \{ \dot{A}_{S}(\vec{w}) \delta ( A_{S}(\vec{r}) - z ) \} \\
        \Tilde{\varepsilon}_{\dot{A}^{2}} \equiv \underset{k \in \mathbb{R}}{\bigcup} \{ ( \dot{A}_{S}^{2}(\vec{w}) - \langle \dot{A}_{S}^{2}(\hat{\vec{w}}) \rangle_{z} ) \delta ( A_{S}(\vec{r}) - z ) \}
    \end{array}
    \right..
\end{align}
Fourth, we use eq.~\ref{si_3_def:projection_operator_decomposition} to decompose $\mathcal{P}_{KE}$ into
\begin{align}
    \label{si_5_eq:p_h_z_z}
    & \mathcal{P}_{KE} O(t, \vec{w}) = \mathcal{P}_{1} O(t, \vec{w}) + \mathcal{P}_{\dot{A}} O(t, \vec{w}) + \mathcal{P}_{\dot{A}^{2}} O(t, \vec{w})
\end{align}
where we already know $\mathcal{P}_{1}$ and $\mathcal{P}_{\dot{A}}$ in eq.~\ref{si_5_eq:p1_2} and
eq.~\ref{si_5_eq:pdotA_2}. Therefore, we compute
\begin{align}
    \label{si_6_eq:inverse_gram_operator}
    & \hat{\mathcal{G}}_{\dot{A}^{2}, \dot{A}^{2}}^{-1}(z, z') = \frac{\delta(z - z')}{\langle  ( \dot{A}_{S}^{2}(\hat{\vec{w}}) - \langle \dot{A}_{S}^{2}(\Tilde{\vec{w}}) \rangle_{z} )^{2} \delta ( A_{S}(\hat{\vec{r}}) - z' ) \rangle},
\end{align}
plug the last result into eq.~\ref{si_3_def:projection_operator_decomposition}, where we substituted
$ ( \dot{A}_{S}^{2}(\vec{w}) - \langle \dot{A}_{S}^{2}(\hat{\vec{w}}) \rangle_{z} ) \delta (
A_{S}(\vec{r}) - z )$ for $E(z, \vec{w})$, and obtain
\begin{align}
    \label{si_6_eq:pdotA2}
    & \mathcal{P}_{\dot{A}^{2}} O(t, \vec{w}) = \frac{\langle O(t, \hat{\vec{w}}) ( \dot{A}_{S}^{2}(\hat{\vec{w}}) - \langle \dot{A}_{S}^{2}(\Tilde{\vec{w}}) \rangle_{A_{S}(\vec{r})} ) \rangle_{A_{S}(\vec{r})}}{\langle ( \dot{A}_{S}^{2}(\hat{\vec{w}}) - \langle \dot{A}_{S}^{2}(\Tilde{\vec{w}}) \rangle_{A_{S}(\vec{r})} )^{2} \rangle_{A_{S}(\vec{r})}} \left( \dot{A}_{S}^{2}(\hat{\vec{w}}) - \langle \dot{A}_{S}^{2}(\Tilde{\vec{w}}) \rangle_{A_{S}(\vec{r})} \right).
\end{align}
Fifth, we insert eq.~\ref{si_6_eq:pdotA2} in eq.~\ref{si_5_eq:p_h_z_z} and recover
eq.~\ref{s2_ss2_def:hybrid_projection_z_z}. We know from
sec.~\ref{si_2:projection_formalism_discrete} that $\mathcal{P}_{KE}$ is a linear, idempotent, and
self-adjoint operator, an orthogonal projection, and has three invariants: $\delta ( A_{S}(\vec{r})
- z )$, $ \dot{A}_{S}(\vec{w}) \delta ( A_{S}(\vec{r}) - z )$, and $ \dot{A}_{S}^{2}(\vec{w}) \delta
( A_{S}(\vec{r}) - z )$. Hence, the observable
% TODO should this be f(k)+g(k) in the integral?
\begin{align}
    \label{si_6_eq:invariant_quantities_1}
    & f(A_{S}(\vec{r})) + \dot{A}_{S}(\vec{w}) g(A_{S}(\vec{r})) + \dot{A}_{S}^{2}(\vec{w}) h(A_{S}(\vec{r})) = \int_{- \infty}^{+ \infty} dz \: \left( f(k) + g(k) \dot{A}_{S}(\vec{w}) + h(k) \dot{A}_{S}^{2}(\vec{w}) \right) \delta ( A_{S}(\vec{r}) - z )
\end{align}
is also an invariant of $\mathcal{P}_{KE}$.

\section{\label{si_7:markvoian_force_1}Derivation of $M_{\text{eff}}(A(t, \vec{w}))$ and $U_{\text{eff}}(A(t, \vec{w}))$}

We derive the expression of the effective force for the I-GLE and D-GLE in
eq.~\ref{s2_ss3_eq:gle_ni_mk} and eq.~\ref{s2_ss3_eq:gle_ni_zk}. First, we recall
eq.~\ref{s2_ss1_def:effective_force} and substitute $\mathcal{P}_{LV}$ for $\mathcal{P}_{1}$ to
obtain
\begin{align}
    \label{si_7_eq:effective_force}
    & F_{\text{eff}}(t, \vec{w}) = e^{t \mathcal{L}} \mathcal{P}_{LV} \ddot{A}_{S}(\vec{w})
\end{align}
where
\begin{align}
    \label{si_7_eq:initial_effective_force}
    & \mathcal{P}_{LV} \ddot{A}_{S}(\vec{w}) = \langle \ddot{A}_{S}(\hat{\vec{w}}) \rangle_{A_{S}(\vec{r})} + \frac{\langle \ddot{A}_{S}(\hat{\vec{w}}) \dot{A}_{S}(\hat{\vec{w}}) \rangle_{A_{S}(\vec{r})}}{\langle \dot{A}_{S}^{2}(\hat{\vec{w}}) \rangle_{A_{S}(\vec{r})}} \dot{A}_{S}(\vec{w}).
\end{align}
Second, we use eq.~\ref{s2_ss2_def:conditional_expectation}, that $\mathcal{L}$ defines an
anti-self-adjoint operator, that $\ddot{A}_{S}(\vec{w}) = \mathcal{L}\dot{A}_{S}(\vec{w})$, and
that~\cite{ayaz_generalized_2022}
\begin{align}
    \label{si_7_eq:liouville_operator_dirac}
    & \mathcal{L} \delta ( A_{S}(\vec{r}) - A ) = - \dot{A}_{S}(\vec{w}) \frac{\partial \delta}{\partial A} ( A_{S}(\vec{r}) - A )
\end{align}
to compute
\begin{align}
    \label{si_7_eq:conditional_expectation_initial_acceleration_2}
    & \langle \ddot{A}_{S}(\hat{\vec{w}}) \rangle_{A} = \frac{1}{\langle \delta ( A_{S}(\hat{\vec{r}}) - A ) \rangle} \frac{d( \langle \delta ( A_{S}(\hat{\vec{r}}) - A ) \rangle \langle \dot{A}_{S}^{2}(\hat{\vec{w}}) \rangle_{A} )}{dA}
\end{align}
which is rearranged into
\begin{align}
    \label{si_7_eq:conditional_expectation_initial_acceleration_3}
    & \langle \ddot{A}_{S}(\hat{\vec{w}}) \rangle_{A} = \langle \dot{A}_{S}^{2}(\hat{\vec{w}}) \rangle_{A} \frac{d \log ( \langle \delta ( A_{S}(\hat{\vec{r}}) - A ) \rangle \langle \dot{A}_{S}^{2}(\hat{\vec{w}}) \rangle_{A} )}{dA}.
\end{align}
Therefore, we recover the definition of $M_{\text{eff}}(A)$ in eq.~\ref{s2_ss3_def:effective_mass}
and the definition of $U_{\text{eff}}(A)$ in eq.~\ref{s2_ss3_def:effective_potential}. Hence,
\begin{align}
    \label{si_7_eq:conditional_expectation_initial_acceleration_4}
    & \langle \ddot{A}_{S}(\hat{\vec{w}}) \rangle_{A} = - \frac{1}{M_{\text{eff}}(A)} \frac{d U_{\text{eff}}(A)}{dA}.
\end{align}
Third, we evaluate $\langle \ddot{A}_{S}(\vec{w}) \dot{A}_{S}(\vec{w}) \rangle_{A_{S}(\vec{r})}$. We
use that $\mathcal{L}$ defines an anti-self-adjoint operator, $\ddot{A}_{S}(\vec{w})
\dot{A}_{S}(\vec{w}) = \mathcal{L} \left( \frac{\dot{A}_{S}^{2}(\vec{w})}{2}  \right)$, and
eq.~\ref{si_7_eq:liouville_operator_dirac} to obtain
\begin{align}
    \label{si_7_eq:two_point_correlation_2}
    & \langle \ddot{A}_{S}(\hat{\vec{w}}) \dot{A}_{S}(\hat{\vec{w}}) \rangle_{A} = \frac{1}{2 \langle \delta ( A_{S}(\hat{\vec{r}}) - A ) \rangle} \frac{d \langle \dot{A}_{S}^{3}(\hat{\vec{w}}) \delta ( A_{S}(\hat{\vec{r}}) - A ) \rangle}{d A}.
\end{align}
Fourth, we recall eq.~\ref{si_6_eq:gram_operator_sixth_block} where $\langle \dot{A}_{S}^{3}(\hat{\vec{w}}) \delta ( A_{S}(\hat{\vec{r}}) - z ) \rangle = 0$ and conclude that $\langle \ddot{A}_{S}(\hat{\vec{w}}) \dot{A}_{S}(\hat{\vec{w}}) \rangle_{A_{S}(\vec{r})}$ vanishes as well. Thus, we obtain
\begin{align}
    \label{si_7_eq:initial_effective_force_2}
    & \mathcal{P}_{LV} \ddot{A}_{S}(\vec{w}) = - \frac{1}{M_{\text{eff}}(A_{S}(\vec{r}))} \frac{d U_{\text{eff}}(A_{S}(\vec{r}))}{dA}
\end{align}
which we plug into eq.~\ref{si_7_eq:effective_force} to derive
\begin{align}
    \label{si_7_eq:effective_force_2}
    & F_{\text{eff}}(t, \vec{w}) = - \frac{1}{M_{\text{eff}}(A(t, \vec{w}))} \frac{d U_{\text{eff}}(A(t, \vec{w}))}{d A}.
\end{align}

\section{\label{si_8:markvoian_force_2}Derivation of $M_{\text{eff}}(A(t, \vec{w}))$ and $U_{\text{eff}}^{KE}(A(t, \vec{w}), \dot{A}(t, \vec{w}))$}

We derive the expression of $F_{\text{eff}}(t, \vec{w})$ for the IKE-GLE and DKE-GLE in
eq.~\ref{s2_ss3_eq:gle_i_mk} and eq.~\ref{s2_ss3_eq:gle_i_zk}. First, we use
eq.~\ref{s2_ss1_def:effective_force} where we substitute $\mathcal{P}_{KE}$ for $\mathcal{P}_{1}$ and
obtain
\begin{align}
    \label{si_8_eq:effective_force}
    & F_{\text{eff}}(t, \vec{w}) = e^{t \mathcal{L}} \mathcal{P}_{KE} \ddot{A}_{S}(\vec{w}),
\end{align}
where
\begin{align}
    \label{si_8_eq:initial_effective_force}
    \mathcal{P}_{KE} \ddot{A}_{S}(\vec{w}) & = \langle \ddot{A}_{S}(\hat{\vec{w}}) \rangle_{A_{S}(\vec{r})} + \frac{\langle \ddot{A}_{S}(\hat{\vec{w}}) \dot{A}_{S}(\hat{\vec{w}}) \rangle_{A_{S}(\vec{r})}}{\langle \dot{A}_{S}^{2}(\hat{\vec{w}}) \rangle_{A_{S}(\vec{r})}} \dot{A}_{S}(\vec{w}) \\
    & + \frac{\langle \ddot{A}_{S}(\hat{\vec{w}}) ( \dot{A}_{S}^{2}(\hat{\vec{w}}) - \langle \dot{A}_{S}^{2}(\Tilde{\vec{w}}) \rangle_{A_{S}(\vec{r})} ) \rangle_{A_{S}(\vec{r})}}{\langle ( \dot{A}_{S}^{2}(\hat{\vec{w}}) - \langle \dot{A}_{S}^{2}(\Tilde{\vec{w}}) \rangle_{A_{S}(\vec{r})} )^{2} \rangle_{A_{S}(\vec{r})}} \left( \dot{A}_{S}^{2}(\vec{w}) - \langle \dot{A}_{S}^{2}(\hat{\vec{w}}) \rangle_{A_{S}(\vec{r})} \right). \nonumber
\end{align}

Second, we know from sec.~\ref{si_7:markvoian_force_1} the expression for $\langle
\ddot{A}_{S}(\hat{\vec{w}}) \rangle_{A_{S}(\vec{r})}$ in
eq.~\ref{si_7_eq:conditional_expectation_initial_acceleration_3} and $\langle
\ddot{A}_{S}(\hat{\vec{w}}) \dot{A}_{S}(\hat{\vec{w}}) \rangle_{A_{S}(\vec{r})} = 0$. Thus, we use
that $\mathcal{L}$ defines an anti-self-adjoint operator,
eq.~\ref{si_7_eq:liouville_operator_dirac}, and that $\ddot{A}_{S}(\hat{\vec{w}})
\dot{A}_{S}^{2}(\hat{\vec{w}}) =  \mathcal{L} \left( \frac{\dot{A}_{S}^{3}(\hat{\vec{w}})}{3}
\right)$ to obtain
\begin{align}
    \label{si_8_eq:third_term_prefactor}
    & \langle \ddot{A}_{S}(\hat{\vec{w}}) \dot{A}_{S}^{2}(\hat{\vec{w}}) \rangle_{A} = \frac{1}{3 \langle \delta ( A_{S}(\hat{\vec{r}}) - A ) \rangle} \frac{d ( \langle \delta ( A_{S}(\hat{\vec{r}}) - A ) \rangle \langle \dot{A}_{S}^{4}(\hat{\vec{w}}) \rangle_{A} )}{dA}.
\end{align}
Third, we use eq.~\ref{si_7_eq:conditional_expectation_initial_acceleration_3},
eq.~\ref{si_8_eq:third_term_prefactor}, and that $\langle \ddot{A}_{S}(\hat{\vec{w}}) \langle
\dot{A}_{S}^{2}(\Tilde{\vec{w}}) \rangle_{A} \rangle_{A} = \langle \ddot{A}_{S}(\hat{\vec{w}})
\rangle_{A} \langle \dot{A}_{S}^{2}(\hat{\vec{w}}) \rangle_{A}$ to derive
\begin{align}
    \label{si_8_eq:initial_effective_force_2}
    & \mathcal{P}_{KE} \ddot{A}_{S}(\vec{w}) = \langle \dot{A}_{S}^{2}(\hat{\vec{w}}) \rangle_{A_{S}(\vec{r})} \frac{d \log \langle \delta ( A_{S}(\hat{\vec{r}}) - A_{S}(\vec{r}) ) \rangle}{dA} + \frac{d \langle \dot{A}_{S}^{2}(\hat{\vec{w}}) \rangle_{A_{S}(\vec{r})}}{dA} + \frac{\dot{A}_{S}^{2}(\hat{\vec{w}}) - \langle \dot{A}_{S}^{2}(\Tilde{\vec{w}}) \rangle_{A_{S}(\vec{r})}}{\langle \dot{A}_{S}^{4}(\hat{\vec{w}}) \rangle_{A_{S}(\vec{r})} - ( \langle \dot{A}_{S}^{2}(\Tilde{\vec{w}}) \rangle_{A_{S}(\vec{r})} )^{2}} \\
    & \left\lbrack \left( \frac{\langle \dot{A}_{S}^{4}(\hat{\vec{w}}) \rangle_{A_{S}(\vec{r})}}{3} - ( \langle \dot{A}_{S}^{2}(\hat{\vec{w}}) \rangle_{A_{S}(\vec{r})} )^{2} \right)  \frac{d \log \langle \delta ( A_{S}(\hat{\vec{r}}) - A_{S}(\vec{r}) ) \rangle}{dA} + \frac{1}{3} \frac{d \langle \dot{A}_{S}^{4}(\hat{\vec{w}}) \rangle_{A_{S}(\vec{r})}}{dA} - \langle \dot{A}_{S}^{2}(\hat{\vec{w}}) \rangle_{A_{S}(\vec{r})} \frac{d \langle \dot{A}_{S}^{2}(\hat{\vec{w}}) \rangle_{A_{S}(\vec{r})}}{dA} \rangle \right\rbrack. \nonumber
\end{align}
We regroup the terms proportional to $\dot{A}_{S}^{2}(\vec{w})$ and obtain
\begin{align}
    \label{si_8_eq:initial_effective_force_3}
    & \mathcal{P}_{KE} \ddot{A}_{S}(\vec{w}) =  - \frac{1}{M_{\text{eff}}(A_{S}(\vec{r}))} \frac{\partial U_{\text{eff}}^{KE}(A_{S}(\vec{r}), \dot{A}_{S}(\vec{w}))}{\partial A}
\end{align}
where we defined the effective potential
\begin{align}
    \label{si_8_def:effective_potential_2}
    - \frac{1}{M_{\text{eff}}(A_{S}(\vec{r}))} \frac{\partial U_{\text{eff}}^{KE}(A_{S}(\vec{w}), \dot{A}_{S}(\vec{w}))}{\partial A} & \equiv \frac{\langle \dot{A}_{S}^{4}(\hat{\vec{w}}) \rangle_{A_{S}(\vec{r})} \frac{d \langle \dot{A}_{S}^{2}(\hat{\vec{w}}) \rangle_{A_{S}(\vec{r})}}{dA} - \frac{1}{3} \langle \dot{A}_{S}^{2}(\hat{\vec{w}}) \rangle_{A_{S}(\vec{r})} \frac{d\langle \dot{A}_{S}^{4}(\hat{\vec{w}}) \rangle_{A_{S}(\vec{r})}}{dA}}{\langle \dot{A}_{S}^{4}(\hat{\vec{w}}) \rangle_{A_{S}(\vec{r})} - ( \langle \dot{A}_{S}^{2}(\hat{\vec{w}}) \rangle_{A_{S}(\vec{r})} )^{2}} \\
    & + \frac{\frac{2}{3} \langle \dot{A}_{S}^{4}(\hat{\vec{w}}) \rangle_{A_{S}(\vec{r})} \langle \dot{A}_{S}^{2}(\hat{\vec{w}}) \rangle_{A_{S}(\vec{r})}}{\langle \dot{A}_{S}^{4}(\hat{\vec{w}}) \rangle_{A_{S}(\vec{r})} - ( \langle \dot{A}_{S}^{2}(\hat{\vec{w}}) \rangle_{A_{S}(\vec{r})} )^{2}} \frac{d \log \langle \delta ( A_{S}(\hat{\vec{r}}) - A_{S}(\vec{r}) ) \rangle}{dA} \nonumber \\
    &  - \frac{1}{2 M_{\text{eff}}(A_{S}(\vec{r}))} \frac{d M_{\text{eff}}(A_{S}(\vec{r}))}{dA} \dot{A}_{S}^{2}(\vec{w}) \nonumber
\end{align}
and the effective mass
\begin{align}
    \label{si_8_def:effective_mass_2}
    - \frac{1}{2 M_{\text{eff}}(A_{S}(\vec{r}))} \frac{d M_{\text{eff}}(A_{S}(\vec{r}))}{dA} & \equiv \frac{\frac{1}{3}\langle \dot{A}_{S}^{4}(\hat{\vec{w}}) \rangle_{A_{S}(\vec{r})} - ( \langle \dot{A}_{S}^{2}(\hat{\vec{w}}) \rangle_{A_{S}(\vec{r})} )^{2}}{\langle \dot{A}_{S}^{4}(\hat{\vec{w}}) \rangle_{A_{S}(\vec{r})} - ( \langle \dot{A}_{S}^{2}(\hat{\vec{w}}) \rangle_{A_{S}(\vec{r})} )^{2}} \frac{d \log \langle \delta ( A_{S}(\hat{\vec{r}}) - A_{S}(\vec{r}) ) \rangle}{dA} \\
    & + \frac{\frac{1}{3} \frac{d \langle \dot{A}_{S}^{4}(\hat{\vec{w}}) \rangle_{A_{S}(\vec{r})}}{dA} - \langle \dot{A}_{S}^{2}(\hat{\vec{w}}) \rangle_{A_{S}(\vec{r})} \frac{d \langle \dot{A}_{S}^{2}(\hat{\vec{w}}) \rangle_{A_{S}(\vec{r})}}{dA}}{\langle \dot{A}_{S}^{4}(\hat{\vec{w}}) \rangle_{A_{S}(\vec{r})} - ( \langle \dot{A}_{S}^{2}(\hat{\vec{w}}) \rangle_{A_{S}(\vec{r})} )^{2}}. \nonumber
\end{align}
Fourth, we consider observables of interest $A(t, \vec{w})$ which satisfy
eq~\ref{s2_ss3_eq:wick_theorem} and recover from eq.~\ref{si_8_def:effective_mass_2}
eq.~\ref{s2_ss3_def:effective_mass} and eq.~\ref{s2_ss3_def:effective_potential_kinetic_energy}. Hence, we insert
\begin{align}
    \label{si_8_eq:initial_effective_force_4}
    & \mathcal{P}_{KE} \ddot{A}_{S}(\vec{w}) = - \frac{1}{M_{\text{eff}}(A_{S}(\vec{w}))} \frac{\partial U_{\text{eff}}^{KE}(A_{S}(\vec{w}), \dot{A}_{S}(\vec{w}))}{\partial A}
\end{align}
into eq.~\ref{si_8_eq:effective_force} and obtain
\begin{align}
    \label{si_8_eq:effective_force_2}
    & F_{\text{eff}}(t, \vec{w}) = - \frac{1}{M_{\text{eff}}(A(t, \vec{w}))} \frac{\partial U_{\text{eff}}^{KE}(A(t, \vec{w}), \dot{A}(t, \vec{w}))}{\partial A}.
\end{align}

\section{\label{si_9:kernel_1}Properties of $F_{\alpha}(t, w)$ and $\Gamma_{\alpha}(t)$}

We derive the properties of $F_{\alpha}(t, w)$ in eq.~\ref{s2_ss3_eq:gle_ni_mk} and
$\Gamma_{\alpha}(t)$ in eq.~\ref{s2_ss3_def:kernel_1}. First, we use
eq.~\ref{s2_ss1_def:orthogonal_force} where we substituted $\mathcal{P}_{M}$ for $\mathcal{P}_{2}$
and $\mathcal{P}_{LV}$ or $\mathcal{P}_{KE}$ for $\mathcal{P}_{1}$. We obtain
\begin{align}
    \label{si_9_eq:orthogonal_force}
    & F_{\alpha}(t, w) = \left( \sum_{n \geq 0} \frac{t^{n}}{n!} (\mathcal{Q}_{M} \mathcal{L})^{n} \right) \mathcal{Q}_{LV/KE} \ddot{A}_{S}(w).
\end{align}
Second, we know on one hand from sec.~\ref{si_4:build_m} that $1$, $A_{S}(\vec{r})$, and
$\dot{A}_{S}(\vec{w})$ are invariants of $\mathcal{P}_{M}$, which implies that
\begin{align}
    \label{si_9_eq:properties_qmnl}
    \left\{
    \begin{array}{c}
        \langle (\mathcal{Q}_{M} \mathcal{L})^{n} \mathcal{Q}_{LV} \ddot{A}_{S}(\hat{\vec{w}}) \rangle = 0 \\
        \langle A_{S}(\hat{\vec{r}}) (\mathcal{Q}_{M} \mathcal{L})^{n} \mathcal{Q}_{LV} \ddot{A}_{S}(\hat{\vec{w}}) \rangle = 0 \\
        \langle \dot{A}_{S}(\hat{\vec{w}}) (\mathcal{Q}_{M} \mathcal{L})^{n} \mathcal{Q}_{LV} \ddot{A}_{S}(\hat{\vec{w}}) \rangle = 0
    \end{array}
    \right.
\end{align}
for all $n \geq 1$. On the other hand, we know from sec.~\ref{si_5:build_h_z} and
sec.~\ref{si_6:build_h_z_z} that $f(A_{S}(\vec{r})) + g(A_{S}(\vec{r})) \dot{A}_{S}(\vec{w})$ is an
invariant of $\mathcal{P}_{LV}$ and $\mathcal{P}_{KE}$. This implies that
\begin{align}
    \label{si_9_eq:properties_qlv}
    & \langle ( f(A_{S}(\hat{\vec{r}})) + g(A_{S}(\hat{\vec{r}})) \dot{A}_{S}(\hat{\vec{w}}) ) \mathcal{Q}_{LV} \ddot{A}_{S}(\hat{\vec{w}}) \rangle = 0 
\end{align}
for arbitrary $f(A)$ and $g(A)$, and in particular for the cases $f(A) = 1$ and $g(A) = 0$, $f(A) =
A$ and $g(A) = 0$, and $f(A) = 0$ and $g(A) = 1$, from which we we recover
eq.~\ref{s2_ss3_eq:orthogonal_force_properties} for $\alpha = L, LKE$. Third, we use
eq.~\ref{s2_ss1_def:memory_kernel} where we substitute $\mathcal{P}_{LV}$ or $\mathcal{P}_{KE}$ for
$\mathcal{P}_{1}$ and $\mathcal{P}_{M}$ for $\mathcal{P}_{2}$ and obtain
\begin{align}
    \label{si_9_eq:memory_kernel}
    & \Gamma(s, t, \vec{w}) = \exp ( s\mathcal{L} ) \mathcal{P}_{M} \mathcal{L} F_{\alpha}(t, \vec{w})
\end{align}
where
\begin{align}
    \label{si_9_eq:memory_kernel_2}
    & \mathcal{P}_{M} \mathcal{L} F_{\alpha}(t, \vec{w}) = \langle \mathcal{L} F_{\alpha}(t, \hat{\vec{w}}) \rangle + \frac{\langle \mathcal{L} F_{\alpha}(t, \hat{\vec{w}}) ( A_{S}(\hat{\vec{r}}) - \langle A_{S}(\Tilde{\vec{r}}) \rangle ) \rangle}{\langle ( A_{S}(\hat{\vec{r}}) - \langle A_{S}(\Tilde{\vec{r}}) \rangle )^{2} \rangle} ( A_{S}(\hat{\vec{r}}) - \langle A_{S}(\Tilde{\vec{r}}) \rangle ) + \frac{\langle \mathcal{L} F_{\alpha}(t, \hat{\vec{w}}) \dot{A}_{S}(\hat{\vec{w}}) \rangle}{\langle \dot{A}_{S}^{2}(\hat{\vec{w}}) \rangle} \dot{A}_{S}(w).
\end{align}
We use that $\mathcal{L}$ defines an anti-self-adjoint operator, that $\mathcal{L}
\rho_{eq}(\vec{w}) = 0$, and eq.~\ref{s2_ss3_eq:orthogonal_force_properties} to obtain
\begin{align}
    \label{si_9_eq:memory_kernel_3}
    & \Gamma(s, t, \vec{w}) = - \Gamma_{\alpha}(t) \dot{A}(s, \vec{w})
\end{align}
where we recover the definition of $\Gamma_{\alpha}(t)$ eq.~\ref{s2_ss3_def:kernel_1}. Third, we recall that $\dot{A}_{S}(\vec{w}) = - \frac{1}{M_{\text{eff}}(A_{S}(\vec{r}))} \frac{d U_{\text{eff}}(A_{S}(\vec{r}))}{dA} + F_{\alpha}(0, \vec{w})$, which we insert into eq.~\ref{s2_ss3_def:kernel_1} to obtain
\begin{align}
    \label{si_9_eq:smr_relation}
    & \Gamma_{\alpha}(t) = \frac{\langle F_{\alpha}(0, \hat{\vec{w}}) F_{\alpha}(t, \hat{\vec{w}}) \rangle}{\langle \dot{A}_{S}^{2}(\hat{\vec{w}}) \rangle} - \frac{1}{\langle \dot{A}_{S}^{2}(\hat{\vec{w}}) \rangle} \left\langle \frac{1}{M_{\text{eff}}(A_{S}(\hat{\vec{r}}))} \frac{d U_{\text{eff}}(A_{S}(\hat{\vec{r}}))}{dA} F_{\alpha}(t, \hat{\vec{w}}) \right\rangle.
\end{align}
We use eq.~\ref{si_5_eq:invariant_observable_1} where we substitute $\frac{1}{M_{\text{eff}}(z)} \frac{d U_{\text{eff}}(A))}{dA}$ for $f(A)$ and $0$ for $g(A)$ with eq.~\ref{s2_ss2_def:conditional_expectation} to rewrite eq.~\ref{si_9_eq:smr_relation} as
\begin{align}
    \label{si_9_eq:smr_relation_2}
    & \Gamma_{\alpha}(t) = \frac{\langle F_{\alpha}(0, \hat{\vec{w}}) F_{\alpha}(t, \hat{\vec{w}}) \rangle}{\langle \dot{A}_{S}^{2}(\hat{\vec{w}}) \rangle} - \frac{1}{\langle \dot{A}_{S}^{2}(\hat{\vec{w}}) \rangle} \int_{-\infty}^{+ \infty} dA \: \frac{1}{M_{\text{eff}}(A)} \frac{d U_{\text{eff}}(A)}{dA} \langle \delta( A_{S}(\hat{\vec{r}} - A) \rangle \left\langle  F_{\alpha}(t, \hat{\vec{w}}) \right\rangle_{A}.
\end{align}
If one demands the second term on the r.h.s. of eq.~\ref{si_9_eq:smr_relation} to vanish irregardless of the choice of $U_{\text{eff}}(A)$ and $M_{\text{eff}}(A)$, one needs $F_{\alpha}(t, \vec{w})$ to satisfy $\left\langle  F_{\alpha}(t, \hat{\vec{w}}) \right\rangle_{A} = 0$. In this case, one recovers eq.~\ref{s2_ss3_eq:SMR_relation}.

\section{\label{si_10:kernel_2}Properties of $F_{\beta}(t, \vec{w})$ and $\Gamma_{\beta}(t, A(s, \vec{w}))$}

We derive the properties of $F_{\beta}(t, \vec{w})$ in eq.~\ref{s2_ss3_eq:orthogonal_force_properties_2} and $\Gamma_{\beta}(t, A(s, \vec{w}))$ in eq.~\ref{s2_ss3_def:kernel_2}. First, we use eq.~\ref{s2_ss3_eq:gle_i_zk}, where we substitute $\mathcal{P}_{KE}$ or $\mathcal{P}_{LV}$ for $\mathcal{P}_{2}$ and $\mathcal{P}_{LV}$ for $\mathcal{P}_{1}$, i.e.
\begin{align}
    \label{si_10_eq:orthogonal_force}
    & F_{\beta}(t, \vec{w}) = \left( \sum_{n \geq 0} \frac{t^{n}}{n!} (\mathcal{Q}_{LV} \mathcal{L})^{n} \right) \mathcal{Q}_{LV/KE} \ddot{A}_{S}(\vec{w}).
\end{align}
Second, we use from sec.~\ref{si_5:build_h_z} that $\mathcal{P}_{LV}$ defines an orthogonal
projection and that any observable of the form $f(A_{S}(\vec{r})) + g(A_{S}(\vec{r}))
\dot{A}_{S}(\vec{w})$ is an invariant of $\mathcal{P}_{LV}$ to derive
\begin{align}
    \label{si_10_eq:properties_qlvn}
    & \langle ( f(A_{S}(\hat{\vec{r}})) + g(A_{S}(\hat{\vec{r}})) \dot{A}_{S}(\hat{\vec{w}}) )(\mathcal{Q}_{LV} \mathcal{L})^{n} \mathcal{Q}_{KE} \ddot{A}_{S}(\hat{\vec{w}}) \rangle = 0
\end{align}
for any $n \geq 1$. Third, we use from sec.~\ref{si_6:build_h_z_z} that $\mathcal{P}_{KE}$ defines
an orthogonal projection and that any observable of the form $f(A_{S}(\vec{r})) + g(A_{S}(\vec{r}))
\dot{A}_{S}(\vec{w}) + h(A_{S}(\vec{r})) \dot{A}_{S}^{2}(\vec{w}) $ is an invariant of
$\mathcal{P}_{KE}$ to derive
\begin{align}
    \label{si_10_eq:properties_qke}
    & \langle ( f(A_{S}(\hat{\vec{r}})) + g(A_{S}(\hat{\vec{r}})) \dot{A}_{S}(\hat{\vec{w}}) + h(A_{S}(\hat{\vec{r}})) \dot{A}_{S}^{2}(\hat{\vec{w}}) ) \mathcal{Q}_{KE} \ddot{A}_{S}(\hat{\vec{w}}) \rangle = 0
\end{align}
for arbitrary functions $f$, $g$ and $h$ of $A_{S}(\vec{r})$, in particular for $f(A) =
\frac{\delta( A_{S}(\vec{r}) - A )}{\langle \delta ( A_{S}(\vec{r}) - A ) \rangle}$, $g(A) = 0$,
$h(A) = 0$, $f(A) = 0$, $g(A) = \frac{\delta( A_{S}(\vec{r}) - A )}{\langle \delta ( A_{S}(\vec{r})
- A ) \rangle}$, $h(A) = 0$, or $f(A) = 0$, $g(A) = \frac{\delta( A_{S}(\vec{r}) - A )}{\langle
\delta ( A_{S}(\vec{r}) - A ) \rangle}$, $h(A) = 0$. Therefore, we recover
eq.~\ref{s2_ss3_eq:orthogonal_force_properties_2} for $\beta = NL, NLKE$. Fourth we use
eq.~\ref{s2_ss1_def:memory_kernel} and compute
\begin{align}
    \label{si_10_eq:memory_kernel}
    & \Gamma(s, t, \vec{w}) = e^{s \mathcal{L}} \mathcal{P}_{LV} \mathcal{L} F_{\beta}(t, \vec{w})
\end{align}
where
\begin{align}
    \label{si_10_eq:memory_kernel_2}
    & \mathcal{P}_{LV} \mathcal{L} F_{\beta}(t, \vec{w}) = \langle \mathcal{L} F_{\beta}(t, \hat{\vec{w}}) \rangle_{A_{S}(\vec{r})} + \frac{\langle \mathcal{L} F_{\beta}(t, \hat{\vec{w}}) \dot{A}_{S}(\hat{\vec{w}}) \rangle_{A_{S}(\vec{r})}}{\langle \dot{A}_{S}^{2}(\hat{\vec{w}}) \rangle_{A_{S}(\vec{r})}} \dot{A}_{S}(\vec{w}).
\end{align}
We use eq.~\ref{s2_ss3_eq:orthogonal_force_properties_2},
eq.~\ref{si_7_eq:liouville_operator_dirac}, and the fact that $\mathcal{L}$ defines an
anti-self-adjoint operator to obtain
\begin{align}
    \label{si_10_eq:memory_kernel_3}
    \mathcal{P}_{LV} \mathcal{L} F_{\beta}(t, \vec{w}) & = - \frac{\langle F_{\beta}(t, \hat{\vec{w}}) \ddot{A}_{S}(\hat{\vec{w}}) \rangle_{A_{S}(\vec{r})}}{\langle \dot{A}_{S}^{2}(\hat{\vec{w}}) \rangle_{A_{S}(\vec{r})}} \dot{A}_{S}(\vec{w}) \\
    & + \frac{1}{\langle \dot{A}_{S}^{2}(\hat{\vec{w}}) \delta ( A_{S}(\hat{\vec{r}}) - A_{S}(\vec{r}) ) \rangle} \frac{d ( \langle \delta ( A_{S}(\hat{\vec{r}}) - A_{S}(\vec{r}) ) \rangle \langle \dot{A}_{S}^{2}(\hat{\vec{w}}) F_{\beta}(t, \hat{\vec{w}}) \rangle_{A_{S}(\vec{r})} )}{dA} \dot{A}_{S}(\vec{w}) \nonumber
\end{align}
which we insert in eq.~\ref{si_10_eq:memory_kernel} to obtain
\begin{align}
    \label{si_10_eq:memory_kernel_4}
    & \Gamma(s, t, \vec{w}) = - \Gamma_{\beta}(t, A(s, \vec{w})) \dot{A}(s, \vec{w}),
\end{align}
by which we then recover the definition of $\Gamma_{\beta}(t, A(s, \vec{w}))$ in eq.~\ref{s2_ss3_def:kernel_2}. Third, we recall that $\ddot{A}_{S}(\vec{r}) = - \frac{1}{M_{\text{eff}}(A_{S}(\vec{r}))} \frac{\partial U_{\text{eff}}^{KE}(A_{S}(\vec{r}), \dot{A}_{S}(\vec{w}))}{\partial A} + F_{\beta}(0, \vec{w})$. We insert this expression into eq.~\ref{s2_ss3_def:kernel_2} and use eq.~\ref{s2_ss3_eq:orthogonal_force_properties_2} to obtain
\begin{align}
    \label{si_10_eq:smr_relation}
    \Gamma_{\beta}(t, A) = & \frac{\langle F_{\beta}(0, \hat{\vec{w}}) F_{\beta}(t, \hat{\vec{w}}) \rangle_{A}}{\langle \dot{A}_{S}^{2}(\hat{\vec{w}}) \rangle_{A}} \\
    & + \frac{\beta \langle \dot{A}_{S}^{2}(\hat{\vec{w}}) F_{\beta}(t, \hat{\vec{w}}) \rangle_{A}}{\langle \dot{A}_{S}^{2}(\hat{\vec{w}}) \rangle_{A}} \frac{d}{dA} \left( U_{\text{pmf}}(A) - \frac{1}{2\beta} \log M_{\text{eff}}(A) \right) - \frac{1}{\langle \dot{A}_{S}^{2}(\hat{\vec{w}}) \rangle_{A}} \frac{\partial \langle F_{\beta}(t, \hat{\vec{w}}) \dot{A}_{S}^{2}(\hat{\vec{w}}) \rangle_{A}}{\partial A} \nonumber
\end{align}
which is an expression that depends on $\langle \dot{A}_{S}^{2}(\hat{\vec{w}}) F_{\beta}(t, \hat{\vec{w}}) \rangle_{A}$ and its partial derivative with respect to $A$. Thus, one recovers eq.~\ref{s2_ss3_eq:SMR_relation_2} if $\langle \dot{A}_{S}^{2}(\hat{\vec{w}}) F_{\beta}(t, \hat{\vec{w}}) \rangle_{A} = 0$ or if $\langle \dot{A}_{S}^{2}(\hat{\vec{w}}) F_{\beta}(t, \hat{\vec{w}}) \rangle_{A}$ satisfies eq.~\ref{s2_ss3_eq:pde_faa_correlation}.

\section{\label{si_11:cg_boltzmann} Wick's theorem and the joint statistics of $A$ and $\dot{A}$}

% TODO check if youu like this Benjamin, I moved this equation from the main text here since it is referenced twice more in the SI
We show that the statistics of $A$ and $\dot{A}$ follow the joint distribution
\begin{align}
    \label{s2_ss4_def:cg_boltzmann_distribution}
    & \langle \delta ( A_{S}(\hat{\vec{r}}) - A ) \delta ( \dot{A}_{S}(\hat{\vec{w}}) - \dot{A} ) \rangle = \frac{e^{- \beta U_{\text{j}}(A, \dot{A})}}{Z_{\text{CG}}(\beta)}
\end{align}
for an arbitrary potential $U_{\text{j}}(A, \dot{A})$ when eq.~\ref{s2_ss3_eq:wick_theorem} holds.
First, we assume that eq.~\ref{s2_ss3_eq:wick_theorem} holds and use
eq.~\ref{s2_ss3_eq:wick_theorem} and that
\begin{align}
    \label{si_11_eq:cg_boltzmann_1}
    & \left\langle \exp(i \lambda \dot{A}_{S}(\hat{\vec{w}})) \right\rangle_{A} = \sum_{n \geq 0} \frac{(i \lambda)^{n}}{n!} \langle \dot{A}_{S}^{n}(\hat{\vec{w}}) \rangle_{A}
\end{align}
to obtain
\begin{align}
    \label{si_11_eq:cg_boltzmann_2}
    & \left\langle \exp(i \lambda \dot{A}_{S}(\hat{\vec{w}})) \right\rangle_{A} = \exp(- \frac{\lambda^{2}}{2} \langle \dot{A}_{S}^{2}(\hat{\vec{w}}) \rangle_{A}).
\end{align}
Second, we use eq.~\ref{si_11_eq:cg_boltzmann_2} and eq.~\ref{s2_ss2_def:conditional_expectation} to compute
\begin{align}
    \label{si_11_eq:cg_boltzmann_3}
    & \langle \delta ( A_{S}(\hat{\vec{r}}) - A ) \delta ( \dot{A}_{S}(\hat{\vec{w}}) - \dot{A} ) \rangle = \frac{1}{2\pi} \int_{-\infty}^{+\infty} d\lambda \: \langle e^{i \lambda \dot{A}_{S}(\hat{\vec{w}})} \delta ( A_{S}(\hat{\vec{r}}) - A ) \rangle e^{-i \lambda \dot{A}}
\end{align}
which can be rewritten as the Gaussian integral
\begin{align}
    \label{si_11_eq:cg_boltzmann_4}
    & \langle \delta ( A_{S}(\hat{\vec{r}}) - A ) \delta ( \dot{A}_{S}(\hat{\vec{w}}) - \dot{A} ) \rangle = \frac{\langle \delta ( A_{S}(\hat{\vec{r}}) - A ) \rangle}{2\pi} \int_{-\infty}^{+\infty} d\lambda \: e^{- \frac{\lambda^{2}}{2} \langle \dot{A}_{S}^{2}(\hat{\vec{w}}) \rangle_{A}} e^{-i \lambda \dot{A}},
\end{align}
from which we obtain that
\begin{align}
    \label{si_11_eq:cg_boltzmann_5}
    & \langle \delta ( A_{S}(\hat{\vec{r}}) - A ) \delta ( \dot{A}_{S}(\hat{\vec{w}}) - \dot{A} ) \rangle = \frac{\langle \delta ( A_{S}(\hat{\vec{r}}) - A ) \rangle}{ \sqrt{2\pi \langle \dot{A}_{S}^{2}(\hat{\vec{w}}) \rangle_{A}} } \exp(-\frac{\dot{A}^{2}}{2 \langle \dot{A}_{S}^{2}(\hat{\vec{w}}) \rangle_{A}}).
\end{align}
Third, we use the definition of
$M_{\text{eff}}(A)$ in eq.~\ref{s2_ss3_def:effective_mass} and the definition of $U_{\text{eff}}(A,
\dot{A})$ in eq.~\ref{s2_ss3_def:effective_potential_kinetic_energy} and obtain
% TODO check if you like this I just give the content of potentials equality here
$U_{\text{j}}(A, \dot{A}) = U_{\text{eff}}^{KE}(A, \dot{A})$.
% eq.~\ref{s2_ss4_def:potentials_equality}.

\section{\label{si_12:example}Examples of observables that satisfy satisfy Wick's theorem}

We check whether eq.~\ref{s2_ss3_eq:wick_theorem} holds for a Schrödinger-type observable of the form
\begin{align}
    \label{si_12_eq:choice_observable}
    & A_{S}(\vec{r}) = \sum_{\alpha = 1}^{N_{d}} f_{\alpha} \left( D_{\alpha}(\vec{r}) \right),
\end{align}
where
\begin{align}
    \label{si_12_eq:squared_norm_alpha}
    & D_{\alpha}(\vec{r}) \equiv \sum_{i_{\alpha}, j_{\alpha} = 1}^{N} \mu_{i_{\alpha}} \mu_{j_{\alpha}} \sum_{k_{\alpha} = 0}^{2} r_{3i_{\alpha} - k_{\alpha}} r_{3j_{\alpha} - k_{\alpha}}
\end{align}
in the cases when $N_{d} = 1$ and $N_{d} > 1$. First, we compute
% TODO did you mean to use \vec{\vec{w}}? there are several occurences
\begin{align}
    \label{si_12_eq:velociy}
    & \dot{A}_{S}(\vec{\vec{w}}) = 2 \sum_{j=1}^{3N} \frac{p_{l}}{m_{l}} \sum_{\alpha = 1}^{N_{d}} \frac{d f_{\alpha} \left( D_{\alpha}(\vec{r}) \right)}{d D} \sum_{i_{\alpha}, j_{\alpha} = 1}^{N} \mu_{i_{\alpha}} \mu_{j_{\alpha}} \sum_{k_{\alpha} = 0}^{2} r_{3i_{\alpha} - k} \delta_{l, 3j_{\alpha} - k_{\alpha}}
\end{align}
and obtain that
\begin{align}
    \label{si_12_eq:mgf_1}
    & \exp(i \lambda \dot{A}_{S}(\vec{\vec{w}})) = \exp(2 \sum_{j=1}^{3N} \frac{p_{l}}{m_{l}} \sum_{\alpha = 1}^{N_{d}} \frac{d f_{\alpha} \left( D_{\alpha}(\vec{r}) \right)}{d D} \sum_{i_{\alpha}, j_{\alpha} = 1}^{N} \mu_{i_{\alpha}} \mu_{j_{\alpha}} \sum_{k_{\alpha} = 0}^{2} r_{3i_{\alpha} - k} \delta_{l, 3j_{\alpha} - k_{\alpha}}).
\end{align}
Second, we average $e^{i \lambda \dot{A}_{S}(\vec{\vec{w}})}$ over the components of $\vec{p}$ and obtain
\begin{align}
    \label{si_12_eq:mgf_2}
    & \prod_{l = 1}^{3N} \int_{- \infty}^{+ \infty} d\hat{p}_{l} \: \exp(- \frac{\beta}{2} \overset{3N}{\underset{l = 1}{\sum}} \frac{\hat{p}_{l}^{2}}{m_{l}} + i \lambda \dot{A}_{S}(\hat{\vec{w}})) = \\
    & \exp(-\frac{2 \lambda^{2}}{\beta} \overset{N_{d}}{\underset{\alpha, \alpha' = 1}{\sum}} \frac{d f_{\alpha} \left( D_{\alpha}(\hat{\vec{r}}) \right)}{d D} \frac{d f_{\alpha'} \left( D_{\alpha'}(\hat{\vec{r}}) \right)}{d D} \overset{N}{\underset{j_{\alpha} = 1}{\sum}} \frac{\mu_{j_{\alpha}}^{2}}{m_{3j_{\alpha}}} \overset{N}{\underset{i_{\alpha}, i_{\alpha'} = 1}{\sum}} \mu_{i_{\alpha}} \mu_{i_{\alpha'}} \overset{2}{\underset{k_{\alpha} = 0}{\sum}} r_{3i_{\alpha} - k_{\alpha}} r_{3i_{\alpha'} - k_{\alpha}}). \nonumber
\end{align}
Third, we consider the case $N_{d} = 1$ where $A_{S}(\vec{r}) = f_{1} \left( D_{1}(\vec{r})
\right)$, i.e. $D_{1}(\vec{r}) = f_{1}^{-1}(A_{S}(\vec{r}))$, and use eq.~\ref{si_12_eq:mgf_2} to
obtain
\begin{align}
    \label{si_12_eq:mgf_4}
    & \left\langle \exp(i \lambda \dot{A}_{S}(\hat{\vec{\vec{w}}})) \delta( A_{S}(\hat{\vec{r}}) - A ) \right\rangle = \exp(- \frac{2 \lambda^{2}}{\beta} f^{-1}(A) \left( f'(f^{-1}(A)) \right)^{2} \overset{3N}{\underset{j = 1}{\sum}} \frac{\mu_{j}^{2}}{m_{3j}}) \left\langle \delta( A_{S}(\hat{\vec{r}}) - A ) \right\rangle
\end{align}
Then, we use eqs.~\ref{si_11_eq:cg_boltzmann_4}, ~\ref{si_11_eq:cg_boltzmann_5}, and~\ref{s2_ss4_def:cg_boltzmann_distribution} to
recover the expression of $M_{\text{eff}}(A)$ in eq.~\ref{s2_ss4_eq:cg_mass}. Fourth, we consider
the case $N_{d} > 1$ and use eq.~\ref{si_12_eq:mgf_2} to compute
\begin{align}
    \label{si_12_eq:mgf_5}
    & \left\langle \exp(i \lambda \dot{A}_{S}(\hat{\vec{\vec{w}}})) \right\rangle_{A} = \left\langle \exp(-\frac{2 \lambda^{2}}{\beta} \overset{N_{d}}{\underset{\alpha, \alpha' = 1}{\sum}} \frac{d f_{\alpha} \left( D_{\alpha}(\hat{\vec{r}}) \right)}{d D} \frac{d f_{\alpha'} \left( D_{\alpha'}(\hat{\vec{r}}) \right)}{d D} \overset{N}{\underset{j_{\alpha} = 1}{\sum}} \frac{\mu_{j_{\alpha}}^{2}}{m_{3j_{\alpha}}} \overset{N}{\underset{i_{\alpha}, i_{\alpha'} = 1}{\sum}} \mu_{i_{\alpha}} \mu_{i_{\alpha'}} \overset{2}{\underset{k_{\alpha} = 0}{\sum}} r_{3i_{\alpha} - k_{\alpha}} r_{3i_{\alpha'} - k_{\alpha}} ) \right\rangle_{A}
\end{align}
which cannot be simplified. Hence, since we cannot recover an equation similar to
eq.~\ref{si_12_eq:mgf_4}, it follows from sec.~\ref{si_11:cg_boltzmann} that the joint statistics of
$A$ and $\dot{A}$ are not satisfying eq.~\ref{s2_ss4_def:cg_boltzmann_distribution}
% TODO Roland comment: you do not show when eq. 22 holds?

\end{document}